\documentstyle [aps,preprint]{revtex}

\begin{document}

\title{Chaotic Properties of Dilute Two and Three Dimensional Random Lorentz
Gases I: Equilibrium Systems}
\author{H. van Beijeren} 
\address{Institute for Theoretical Physics, University of Utrecht, \\
Postbus 80006, Utrecht 3508 TA, The Netherlands} 
\author{Arnulf Latz and J. R. Dorfman} 
\address{Institute for Physical Science and Technology, and
Department of Physics, \\  
University of Maryland, College Park, Maryland,
20742, USA}
\date{\today}
\maketitle

\begin{abstract}
We compute the Lyapunov
spectrum and the Kolmogorov-Sinai entropy for a moving particle
placed in a dilute, random array of hard disk or hard sphere
scatterers - i.e. the dilute Lorentz gas model. This is carried out in two
ways: First we use simple kinetic theory arguments to compute the Lyapunov
spectrum for both two and three dimensional systems. In order to provide a
method that can easily be generalized to non-uniform systems we then use a
method based upon extensions of the Lorentz-Boltzmann (LB)  equation to
include
variables that characterize the chaotic behavior of the system. The
extended LB equations 
depend upon the
number of dimensions and 
on whether one is computing positive or negative
Lyapunov exponents. In the latter case the extended LB equation is
closely related to an ``anti-Lorentz-Boltzmann equation" where the
collision operator has the opposite sign from the ordinary LB
equation. Finally we compare our results with computer simulations of
Dellago and Posch
and find very good agreement. 
\end{abstract}

\section{Introduction}
The Lorentz model of a gas of non-interacting particles which collide
with fixed scatterers has been a basic model for acquiring an
understanding of fundamental issues in both the kinetic theory of
gases and the general theory of non-equilibrium phenomena in fluids
and solids \cite{Hauge,Cohen1}. In this series of papers we use the
Lorentz gas to study important features of the chaotic behavior of
systems which show typical equilibrium and non-equilibrium behavior
such as the existence of a spatially homogeneous equilibrium state,
and normal diffusion of moving particles among the scatterers. Since
the particles do not collide with each other the Lorentz gas can
readily be analyzed in terms of the properties of one moving particle
in the set of fixed scatterers.  Here we consider the scatterers to be
placed at random in space, subject only to the condition that they are
not allowed to overlap with each other. The chaotic properties of a
particle moving in a {\em periodic} array of non-overlapping hard
disks have been studied extensively, especially for the case where the
density of scatterers is sufficiently high that the moving particle is
unable to travel unimpeded through the lattice (the case of finite
horizon) \cite{Sinai1}. It is known that the random Lorentz gas is a
K-system \cite{Sinai2}, and that the periodic Lorentz gas with finite
horizon is a Bernoulli system \cite{Gal-Or,Che-Has}.  These results are
sufficient to prove that the gas has a well defined equilibrium state,
and that suitably defined initial ensemble distributions will approach
equilibrium distributions for long enough times. However for the
random case there are very few analytic results for quantities that
characterize the chaotic behavior of the moving particle. There is a
conjecture by Krylov that the positive Lyapunov exponents for the
moving particle are proportional to 
$na^{d-1}v\ln[na^{d}]^{-1}$ 
if
$na^d \ll 1$ 
where $n=N/V$ is the number density of $N$ scatterers of
radius $a$ in volume $V$, and $v$ is the constant speed of the moving
particle \cite{Krylov}. This conjecture has been verified by Chernov,
who argued that for low enough densities the periodic and 
the random
Lorentz gas
should have the same value of the Kolmogorov-Sinai
entropy, and then calculated this quantity for a periodic system at
low density. Chernov obtained the results \cite{Chernov}
\begin{eqnarray}
h_{KS}=\lambda^{+} \simeq -2nav \ln na^2 \;\;\; {\rm for}\,\, d=2 \\   
h_{KS}=\lambda^{+}_1 + \lambda^{+}_2 \simeq -2\pi n a^2v \ln na^3 
\;\;\; {\rm for}\,\, d=3.
\label{1}
\end{eqnarray}
Here $\lambda^{+}$ denotes a positive Lyapunov exponent. Simple
considerations of the number of degrees of freedom and the
conservation of energy show that for a two
dimensional Lorentz gas there can 
be 
no more than
one positive Lyapunov exponent,
and for a three dimensional gas there can be at most two of them. The
quantity $h_{KS}$ is the Kolmogorov-Sinai (KS) entropy, which for a
closed, isolated
ergodic
system, such as the one considered by Chernov, is equal to the sum of
the positive Lyapunov exponents, according to Pesin's
theorem\cite{Ru-Eck}. Chernov's results are only the first terms in the 
density expansion of $h_{KS}$ and up till the present work no further
analytic results were known either for the density dependent
corrections to these results or, for three dimensional systems, the
individual Lyapunov exponents. 

We have been able to use familiar methods
from the kinetic theory of gases to calculate the Lyapunov spectrum and
the KS entropy for random Lorentz gases at low
densities\cite{hvb1,hvb2,latz1,latz}. We can do
this for closed, isolated systems as well as for closed systems in a 
magnetic field,
open systems (with
 THESE TWO CATEGORIES ARE THE SAME AS FAR AS I KNOW
escape of particles),
and for systems where the moving particle is
charged and subjected to an electric field plus a thermostat which
maintains a constant kinetic energy even in the presence of the
electric field. These two latter cases are of particular interest
because of their importance for
methods
that relate dynamical
quantities such as Lyapunov exponents and KS entropies to transport
coefficients, in this case the diffusion coefficient for the moving
particle\cite{Gas-Nic,Evans-Hoo}. This series of papers will be
devoted to obtaining the low
density results for Lyapunov exponents and KS entropies under the
various situations mentioned above. Extensions of these results to
higher densities and
to other quantities will be presented elsewhere.

In this paper we shall consider dilute, equilibrium Lorentz gases
in two or three dimensions, consisting, respectively,  of randomly
placed but non-overlapping, fixed hard
disk or hard sphere scatterers, and a point particle of mass $m$ and speed
$v$ moving among them. The collisions with the scatterers are taken to
be elastic. In subsequent papers
we will generalize this model to non-equilibrium situations with
thermostatted electric fields, and/or  open systems with absorbing
boundaries. The plan of this paper is as follows: In Section II we
present an elementary kinetic theory argument which correctly provides
the low density values of the Lyapunov exponents and KS entropy of the
Lorentz gas in two and three dimensions for closed systems and without
external fields. In Section III we consider a more formal approach to these
quantities due to Sinai\cite{Sinai1},
which will form the basis of the extension of
the kinetic theory approach to non-uniform systems. There we
provide the fundamental geometric formulas of Sinai which relate the
Lyapunov exponents and Kolmogorov-Sinai (KS) entropy of a Lorentz gas to
the properties of a radius of curvature matrix. Then, we use the ergodic
properties of the moving particle to express the Lyapunov exponents
and KS entropy in terms of averages over an equilibrium
ensemble. In Section IV we show that the 
pertinent
distribution functions
can be obtained from the solution of an appropriate
extended
Lorentz Boltzmann (LB) equation, and in Section V we calculate the KS entropy of the two and
three dimensional Lorentz gas at low densities.
In Section VI we consider the negative Lyapunov exponents and
show how they can be obtained from a solution of an
``anti-Lorentz-Boltzmann" equation. This will be important for the
extension of the theory to treat thermostatted systems. We conclude in Section VII with a 
comparison of these results
with the results of computer simulations by Posch and Dellago\cite{Pos-Del}, and with a discussion of
the applications of the methods developed here to more general
systems.

\section{Velocity Deviation Method for Lyapunov Exponents and KS Entropies}
 
We consider a system of $N$ $d$-dimensional hard sphere scatterers
placed randomly in space in a $d$-dimensional volume $V$ at low density. Here $d=2,3$, the spheres
have a radius $a$, and the number density of the spheres, $n=N/V$
satisfies $na^{d} \ll 1$. The moving particle travels freely between
elastic collisions with the scatterers. The phase point $x$ of the
particle, i.e. its position and velocity,  $x=({\vec r},{\vec v})$
satisfies the equations
of motion $\dot{{\vec r}}={\vec v};\,\, \dot{{\vec
v}} =0$ between collisions. At a collision of the moving particle
with a scatterer, the velocity of the moving particle changes
according to
\begin{equation}
{\vec v}^+={\vec v}-2({\vec v}\cdot {\hat n}){\hat n},
\label{2}
\end{equation}
where ${\vec v}^+$ is the velocity after collision, and ${\hat n}$ is the
unit vector in the direction from the center of the scatterer to the point of impact of
the moving particle at collision\cite{Chap-Cow}. This completely specifies the
dynamics of the particle given its initial phase and the locations of
all of the scatterers.

The Lyapunov exponents characterize the rate of exponential separation
or of exponential convergence of infinitesimally nearby trajectories on
the $2d-1$ dimensional constant energy surface $\cal{E}$ \cite{Ru-Eck}. 
Since the Lorentz gas is a symplectic,
Hamiltonian system, if there are non-zero
Lyapunov exponents they come in pairs of positive and negative
values, $\pm
\lambda_{i}$\cite{Arnold}. However, the Lyapunov exponent for displacements in the direction of the
trajectory is zero since two phase points on the same physical
trajectory will follow one another without exponential separation or
contraction. Therefore there can be at
most $d-1$ positive and $d-1$ negative Lyapunov exponents for our
system. 
  
To treat the Lyapunov exponents we consider 
a bundle of infinitesimally nearby trajectories on
$\cal{E}$ and follow the motion of this bundle in time.
If the phase point 
on one reference trajectory in this bundle
is 
given
by $x(t)$ we denote the deviation of 
another trajectory in the bundle
from $x(t)$ by $\delta x(t) =(\delta \vec{r}(t), \delta
\vec{v}(t))$. Equations of motion for $\delta x(t)$ follow immediately
from the equations of motion for $x(t)$. 
Since the moving particle has only kinetic energy the requirement that both
trajectories lie on the same energy surface immediately leads to the property
$\vec{v}(t)\cdot\delta\vec{v}(t)=0$ for all $t$.
Without loss of generality, 
we may 
replace $\delta \vec{r}(t)$ by 
the vector of closest
approach between the two trajectories,
i.e. we set $\delta \vec{r}(t) \cdot\vec{v}(t)=0$ 
(from here on we will use
the notation $\delta \vec{r}(t)$ for this vector of closest approach of the
perturbed trajectory to $\vec{r}(t)$.
Notice that if $\delta \vec{r}(t) \cdot\vec{v}(t)=0$ at $t=0$ it remains so at all later times, by virtue of Eqs.\ (3-6).). In between collisions the spatial and velocity deviations
change with time according to
\begin{eqnarray}
\delta\dot{\vec{r}}&=&\delta\vec{v} \label{3a}\\
\delta \dot{\vec{v}} &=& 0.
\label{3}
\end{eqnarray}

The change of $\delta x(t)$ at collisions requires some analysis which
has been provided by Gaspard and Dorfman\cite{Gas-Do} and by Dellago, Posch and
Hoover\cite{DePoHo}. These authors have shown that the change in $\delta x$ at the
instant of a collision is given by
\begin{equation}\label{4a}
\delta \vec{r}^+ = [{\bf 1}-2 \hat{n} \hat{n}]\cdot \delta \vec{r} \\
\end{equation}
\begin{equation}
\delta \vec{v}^+ = [{\bf 1} - 2 \hat{n} \hat{n}]\cdot \delta \vec{v}+
\frac{2}{a} [ \vec{v}\hat{n}
-\hat{n}\vec{v}
+\frac{v^2}{(\vec{v}\cdot\hat{n})}\hat{n}\hat{n}-(\vec{v}\cdot\hat{n}){\bf
1}]\cdot \delta \vec{r}.
\label{4}
\end{equation}
where $\vec{v}, \delta\vec{r}$, $\delta\vec{v}$ are the velocity of
the moving particle, the spatial deviation and the velocity deviation
of the nearby trajectory, immediately before the collision with the
scatterer, while the ``$+$" variables denote the values immediately
after collision. It is important to note here that the
velocity deviation,
 $\delta \vec{v}$, does not change between collisions but does undergo
an instantaneous change at each collision with a scatterer. 

Suppose that we prepare a trajectory bundle with initial
values of $\delta x(0)$ and we follow the motion $\delta x(t)$ in
time. We can relate the largest positive Lyapunov exponent to the
asymptotic growth of the ratio $|\delta \vec{v}(t)|/|\delta
\vec{v}(0)|$, by
\begin{equation}
\lambda_{max}=\lim_{t\rightarrow \infty}
\frac{1}{t}\ln \left[
\frac{|\delta \vec{v}(t)|}{|\delta\vec{v}(0)|}\right].
\label{5}
\end{equation}
This follows from the observation that if there are positive Lyapunov
exponents, then two infinitesimally close trajectories will eventually
separate in time unless they are so precisely arranged that they
approach each other exponentially in time. However the latter
situation occurs only on sets of zero measure (the stable manifold) in 
tangent 
space. Furthermore,
this exponential separation occurs both in configuration space and in
velocity space with the same exponential factor,
since velocities and
positions are related by a simple time derivative which does not
affect the exponential separation rate,
and we may consider the
separation in velocity space alone. In the next section we consider
another 
calculation 
of the Lyapunov exponents,
which treats the separation in
configuration space,
and we obtain the same results.

Similarly, the sum of all of the positive Lyapunov exponents can be
obtained by following the growth of a volume element in velocity space
as
\begin{equation}
\sum_{\lambda _{i}>0}\lambda_{i} = \lim_{t \rightarrow
\infty}
\frac{1}{t}\ln | \det {\bf A}(t)|,
\label{6}
\end{equation}
with ${\bf A}(t)$ describing the linear relationship between 
$\delta \vec{v}(t)$ and $\delta \vec{v}(0)$, i.e.
\begin{equation}
\delta \vec{v}(t)={\bf A}(t)\cdot \delta \vec{v}(0),
\label{7}
\end{equation} 
This result follows because the time evolution of the 
vector $\delta \vec{v}$, which has $d-1$ independent components,
is dominated by the $(d-1)$ largest eigenvalues and
corresponding eigenvectors of the matrix ${\bf A}(t)$, which are precisely
the positive eigenvalues.
Suppose now that
the moving particle undergoes a series of $\cal{N}$ collisions in the time
interval $(0,t)$ with scatterers which we label
$s_{1},s_{2},...,s_{\cal{N}}$. 
Since the width of the
trajectory bundle is 
infinitesimal,
each trajectory within it has the
same number of collisions with each scatterer in the same 
time. 
Since $\delta \vec{v}$, as noted before, changes only at collisions, one has
\begin{equation}
\frac{|\delta \vec{v}(t)|}{|\delta \vec{v}(0)|}=\frac{|\delta
\vec{v}^{+}_{\cal{N}}|}{|\delta \vec{v}^{+}_{{\cal{N}}-1}|}\frac{|\delta
\vec{v}^{+}_{{\cal{N}}-1}|}{|\delta \vec{v}^{+}_{{\cal{N}}-2}|}\cdots\frac{|\delta
\vec{v}^{+}_1|}{|\delta \vec{v}_0|},
\label{9}
\end{equation}
where $\delta\vec{v}^{+}_{i}$ is the velocity deviation immediately
after the collision with scatterer $s_i$. 
For the same reason ${\bf A}(t)$ can change
with time only
at the instants of the collisions of the moving particle with the
scatterer, so that
\begin{equation}
\delta \vec{v}(t) =\delta \vec{v}^{+}_{\cal N} ={\bf a}_{\cal{N}}\cdot
\delta \vec{v}^{+}_{{\cal {N}}-1} ={\bf a}_{\cal N}\cdot{\bf a}_{{\cal
{N}}-1}\cdots{\bf a}_1\cdot \delta \vec{v}(0).
\label{10}
\end{equation}
Here ${\bf a}_j$ is a matrix, to be defined below for the case where
the density of scatterers is low, that expresses the
change in the velocity deviation 
when
the moving particle collides
with scatterer $s_j$. Consequently,
\begin{equation}
\det{\bf A}(t) = \prod_{i}^{\cal{N}}\det{\bf a}_i.
\label{11}
\end{equation}
Of course the number of collisions, ${\cal{N}}$ in time $t$ will depend
upon $t$ and the initial value of the phase point $x(0)$. Expressions
for the largest Lyapunov exponent and for the sum of positive Lyapunov
exponents can be obtained from Eqs. (\ref{5} - \ref{11}) as
\begin{eqnarray}
\lambda_{max}&=&\lim_{t \rightarrow \infty}\frac{{\cal{N}}}{t}\frac{1}{{\cal
{N}}}\sum_{1}^{\cal{N}} \ln\frac{|\delta\vec{v}^{+}_i|}{|\delta\vec{v}^{+}_{i-1}|}
\,\,\, {\rm and}\label{12a} \\
\sum_{\lambda_i >0}\lambda_i &=& \lim_{t \rightarrow \infty}\frac{{\cal
{N}}}{t}\frac{1}{{\cal{N}}} \sum_{1}^{{\cal{N}}} \ln |\det {\bf a}_i| 
\label{12b}
\end{eqnarray}

To proceed further we need to use the fact that the density of
the 
scatterers is small, that is, $na^d \ll 1$. This will allow us to
determine the mean of
the quantities appearing in the sums in Eqs. (\ref{12a}, \ref{12b}). Referring to Eq. 
(\ref{4}) we note that $\delta \vec{r}$ appearing on
the right hand side is the spatial deviation of the moving particle
just before a collision with a scatterer. Let us suppose that we
consider the collision with scatterer $s_i$. Then immediately
before this collision $\delta\vec{r}^-_{i} = \delta
\vec{r}^+_{i-1}+\tau_i\delta \vec{v}^+_{i-1}$, where $\delta
x^+_{i-1}$ 
denotes the spatial and velocity deviations just after the
collision with the previous scatterer, $s_{i-1}$, and $\tau_i$ is the
time between the collision with scatterer $s_{i-1}$ and scatterer
$s_i$. At low scatterer densities the time between collisions will
typically be inversely proportional to the density of scatterers,
so that the ratio of the order of magnitudes of the first to the second term 
in the
above expression for $\delta\vec{r}^-_{i}$ will approach zero as the
scatterer density decreases. Therefore to leading order in the
density, $\delta\vec{r}^-_{i}=\tau_{i}\delta\vec{v}_{i-1}^+$. We then
obtain a low density value for $\delta\vec{v}^+_i$ given by
\begin{equation}
\begin{array}{lcl}
\delta\vec{v}^+_i &=&\left[\left({\bf
1}-2\hat{n}\hat{n}\right)
+\frac{2\tau_i}{a}\left(\vec{v}^+_{i-1}\hat{n} -\hat{n}\vec{v}^{+}_{i-1}-
(\vec{v}^+_{i-1}\cdot\hat{n}){\bf
1}+\frac{v^2}{(\vec{v}^+_{i-1}\cdot\hat{n})}\hat{n}\hat{n}\right)\right]
\cdot\delta\vec{v}^+_{i-1}\\
&
\equiv&{\bf
a}_{i}\cdot\delta\vec{v}^{+}_{i-1}.
\label{13}
\end{array}
\end{equation}  
Now we have expressions for the change in the velocity deviation at
collision and for the matrix ${\bf a}$, both of which are needed for
the calculations outlined above. To evaluate the sums appearing in
Eqs. (\ref{12a}, \ref{12b}), we note that at low densities none of the 
collisions
are correlated with any previous collision, that is, the leading
contribution to the Lyapunov exponents come from collision sequences
where the moving particle does not encounter the same scatterer more
than once in the sequence \footnote{In two dimensions the particle
will hit the same scatterer an infinite number of times. However the
effects of such processes are of higher density, and can be neglected
here since the times between successive collisions with the same
scatterer become
typically
very large as the density of scatterers approaches zero.}. 
Therefore we can treat each term in the
sums in Eqs. (\ref{12a}, \ref{12b}) as being independent of the other terms 
in the
sum. We have expressed $\lambda_{max}$ and the sum of the positive
Lyapunov exponents as arithmetic averages, but for long times and with
independently distributed terms in the average, we can replace the
arithmetic averages by ensemble averages over a suitable equilibrium
ensemble. That is
\begin{equation}
\lambda_{max}=\nu \left<
\ln \left[\frac{|\delta\vec{v}^{+}|}{|\delta\vec{v}_{-}|}\right]\right>
\label{14}
\end{equation}
and
\begin{equation}
\sum_{\lambda_i >0}\lambda_i = \nu \left< \ln |\det{\bf a}|\right>
\label{15}
\end{equation}
where $\nu$ is the (low density)value of the collision frequency,
${\cal N}/t$ as $t$ becomes large, and the angular brackets denote an
equilibrium average.

We now consider a typical collision of the moving particle with one of
the scatterers. The free time between one collision and the next is
sampled from the normalized equilibrium distribution of free times\cite{Chap-Cow}, $P(\tau)$
given at low densities by
\begin{equation}
P(\tau)=\nu e^{-\nu\tau}.
\label{16}
\end{equation}
The construction of the matrix ${\bf a}$ requires some geometry and
depends on the number of dimensions of the system. In any case we take
the velocity vector before collision, $\vec{v}$ to be directed along
the $z$-axis, and take $\hat{n}\cdot\vec{v}=-v\cos\phi$, where $-\pi/2
\leq \phi \leq \pi/2$. The velocity deviation before collision $\delta\vec{v}^-$ is
perpendicular to the $z$-axis. Then it is a simple matter to compute
$|\delta\vec{v}^{+}|/|\delta\vec{v}^-|$ and $|\det{\bf a}|$. For
two-dimensional systems
$\delta \vec{v}$ and
the matrix ${\bf a}$ 
are given in this representation by
\begin{equation}
\delta {\vec{v}}^-=\left(\begin{array}{c}1\\0  \end{array}\right)
|{\delta \vec{v}}^-| \, ; \ \ \ \ 
{\bf a}=\left( \begin{array}{cc}
(1+ \Lambda) \cos 2\phi
 & \sin2\phi
\\
(1+\Lambda) \sin2\phi & 
-\cos2\phi
\end{array}\right),
\label{17}
\end{equation}
where we introduced $\Lambda = (2 v \tau)/(a \cos \phi)$.
To leading order in $v\tau/a$ we find that
\begin{equation}
\frac{|\delta\vec{v}^+|}{|\delta\vec{v}^-|}=
\Lambda; \ \ \ \ \ \ \ \ \ \ \ |\det{\bf a}|
=
\Lambda.
\label{18}
\end{equation}
For three dimensional systems the unit vector $\hat{n}$ can be
represented as $\hat{n}=-\cos\phi \;
\hat{z}+\sin\phi \cos\alpha \; \hat{x} + \sin\phi \sin\alpha \; \hat{y}$.
 Now the 
ranges of the angles $\phi$ and $\alpha$ are $0 \leq
\phi \leq \pi/2$
and 
$0 \leq \alpha \leq 2\pi$. There is an additional angle 
$\psi$ in the $
x,y$ plane such that the 
velocity deviation before collision
$\delta\vec{v}^-=|\delta\vec{v}^-|[\hat{x}\cos\psi +
\hat{y}\sin\psi]$. It is somewhat more convenient to use a symmetric
matrix, $\tilde{\bf{a}} = (\bf{1}-2\hat{n}\hat{n})\cdot\bf{a}$, given
by
\begin{equation}
\tilde{\bf{a}}=
\left( \begin{array}{ccc}
 1+\Lambda(\cos^{2} \phi + \sin^{2} \phi \cos^{2} \alpha ) &
\Lambda \sin^{2} \phi \cos \alpha \sin \alpha & 0 \\
\Lambda \sin^{2} \phi \cos \alpha \sin \alpha & 1 + \Lambda (\cos^{2}
\phi + \sin^{2}\phi \sin^{2} \alpha) & 0 \\
0 & 0 & 1 
\end{array} \right),
\label{18a}
\end{equation}
One easily finds
\begin{equation}
\frac{|\delta\vec{v}^+|}{|\delta\vec{v}^-|}=\frac{2\tau
v}{a}\left[\frac{\cos^2(\alpha-\psi)}{\cos^2\phi}+\sin^2(\alpha-\psi)
\cos^2\phi\right]^{1/2},
\label{19}
\end{equation}
and
\begin{equation}
\det{\tilde{\bf a}}=\det{\bf a}=\left(\frac{2v\tau}{a}\right)^2.
\label{20}
\end{equation}
to leading order in $v\tau/a$. 

To complete the calculation we must evaluate the averages appearing in
Eqs.(\ref{14}, \ref{15}). That is we average over the distribution of 
free times
and over the rate at which scattering events are taking place with the
various scattering angles. Additionally in 3 dimensions an average over a 
stationary distribution of angles $\psi$ has to be performed in general.
 Due to the isotropy of the scattering geometry $\psi$ can here be 
absorbed in a 
redefinition $\alpha' =\alpha -\psi$ of the azimuthal angle 
$\alpha$. This will not be true any
more if the 
isotropy of velocity 
space is
broken (e.g. by an external field). 
The appropriate average of a quantity $F$ takes the simple form
\begin{equation}
\left< F \right> = \frac{
1}{J}\int_0^\infty d\,\tau\int\,
d\hat{n}
\cos\phi P(\tau)F,
\label{21}
\end{equation}
where $P(\tau)$ is the free time distribution given by Eq. (\ref{16})
and $J$ is a normalization factor obtained by setting $F=1$ in the
numerator. The integration over the unit vector $\hat{n}$, i.e.,
over the appropriate solid angle, ranges
over $-\pi/2 \leq \phi \leq \pi/2$ in two dimensions and over 
$0 \leq \phi \leq \pi/2
$ and $
0 \leq \alpha \leq 2\pi$ in three dimensions.
After carrying out the required integrations we find that
\begin{equation}
\lambda^+ = \lambda_{max} = 2nav[-\ln(2na^2) +1 -{\cal C}]+\cdots
\label{22}
\end{equation}
for two dimensions. Here ${\cal C}$ is Euler's constant, and the terms
not given explicitly in Eq. (\ref{22}) are higher order in the
density. Similarly, for the three dimensional Lorentz gas we obtain
\begin{equation} \label{lammax}
\lambda^{+}_{max} = na^{2}v\pi[-\ln(\tilde{n}/2) +\ln 2 -\frac{1}{2}-{\cal
C}]+\cdots,
\end{equation}
\begin{equation}
\lambda^{+}_{max}+\lambda^{+}_{min} = 2na^{2}v\pi[-\ln(\tilde{n}/2) -{\cal C}]+\cdots
\label{23}
\end{equation} 
from which it follows that
\begin{equation}
\lambda^{+}_{min}= na^{2}v\pi[-\ln(\tilde{n}/2) -\ln 2
+\frac{1}{2}-{\cal C}]+\cdots,
\label{24}
\end{equation}
where $\tilde{n}=na^3\pi$. 
We have therefore determined the Lyapunov spectrum for the equilibrium Lorentz
gas
at low densities
in both two and three dimensions 
\cite{hvb1,latz1}. We note that the two positive Lyapunov exponents for three dimensions differ
slightly, and that we were able to get individual values because we
could calculate the largest exponent and the sum of the two exponents.
We could not determine all of the Lyapunov exponents for a $d > 3$
dimensional Lorentz gas this way. Moreover, for a spatially
inhomogeneous system, such as those considered in the application of
escape-rate methods, the simple kinetic arguments used here are not
sufficient and Boltzmann-type methods are essential for the
determination of the Lyapunov exponents and KS entropies. 

We will comment further on these results in Section VII after
we have obtained them again by a more formal method based upon the
radius of curvature matrix method of Sinai\cite{Sinai1}.

\section{The Radius of Curvature Matrix}
Our analysis in this Section is based upon the geometric arguments given by Sinai \cite{Sinai1}
for
the relationships between the Lyapunov exponents, the KS entropy, and the
radius of curvature matrix,
which describes the time evolution of the separation
of nearby phase space trajectories of the moving particle. Unlike the
method presented in the previous section this method treats the time
evolution of the {\em
spatial} separation of a bundle of trajectories rather than the
evolution of the velocity separation of the bundle. Here we
summarize these considerations, referring the reader to the literature
for further details\cite{Sinai1,Gas-Do}. 

As before, the trajectory of the moving particle is specified by the
phase ${x}(t)=(\vec{r}(t),\vec{v}(t))$, and we take a $2d-2$
dimensional 
plane $\Sigma$, $(\delta \vec{r}_{\perp}(t),\delta \vec{v}_{\perp}(t))$, through $ x(t)$, where both
$\delta \vec{ r}_{\perp}(t)$ and $\delta
\vec{ v}_{\perp}(t)$ are perpendicular to the velocity $\vec{v}(t)$. The nearby
trajectories will intersect $\Sigma$, and we measure the separation of
the trajectories by vectors of dimension $2(d-1)$ in $\Sigma$,
$\delta x_{\perp}(t)$, given as
\begin{equation} \label{25}
\delta x_{\perp}(t)= \left(
\begin {array}{c}
\delta \vec{r}_{\perp}(t) \\
\delta \vec{v}_{\perp}(t) 
\end{array} \right ).
\end{equation}
The time development of $\delta x_{\perp}$ is given in terms of a
monodromy matrix 
${\bf M}(t, t_0)$ 
satisfying
\begin{equation} 
\delta  x_{\perp}(t)= {\bf M}(t,t_0) \cdot \delta  x_{\perp}(t_0),
\label{26}
\end{equation} 
The
matrix ${\bf M}
(t,t_0)$ follows the motion of the particle. It changes
continuously with time $t$ in the intervals between collisions, and
undergoes a
discontinuous change at the
instants of collisions of the particle with the scatterers. Between
collisions the monodromy matrix has the form
\begin{equation} \label{27}
{\bf M}(t,t_0)_{\rm{free\,flight}}= \left(
\begin{array}{cc}
{\bf 1} & 
(t-t_0){\bf 1}\\
{\bf 0} & {\bf 1}
\end{array}  \right).
\end{equation}

At the instant of a collision there is a
discontinuous rotation of the velocity of the moving particle from its
value before collision, $\vec{v}$, to its value after
collision, $\vec{v}^+=\vec{v}-2\hat{n}(\vec{v}\cdot\hat{n})$, as before.
Since the velocity of the particle
changes discontinuously at collision, the plane $\Sigma$ also
rotates, and the components of the displacement vector $\delta
x_{\perp}$ change instantaneously. The changes in the components of
$\delta \vec{r}_{\perp}$ and $\delta\vec{v}_{\perp}$ at the
instant of collision are given in Eqs. (\ref{4a}, \ref{4}) above.

In order to determine the Lyapunov exponents for this system we need
to examine the rate of separation or of approach of
infinitesimally close trajectories. This can now be done
with the aid of 
the radius of curvature
operators
$\mbox{\boldmath $\rho$}_{u}$ and $\mbox{\boldmath $\rho$}_{s}$, acting on a
$d-1$ dimensional space of velocity deviation
vectors orthogonal to $\vec{v}$. Here the subscripts $u$ and $s$ denote 
operators describing unstable, or expanding, and stable, or contracting, 
trajectory bundles respectively.
The
operator $\mbox{\boldmath $\rho$}_{u}$ is defined by the relation that
\begin{equation}
\label{28}
 \delta\vec{r}_{\perp}(t) = \frac{1}{v}\mbox{\boldmath $\rho$}_{u}(t)\cdot \delta\vec{v}_{\perp}(t)
,
\end{equation}
together with the conditions $\mbox{\boldmath $\rho$}_{u}
\cdot \vec{v}=\vec{v}\cdot\mbox{\boldmath $\rho$}_{u}=0$.
This definition is motivated by the observation that if the velocity
deviation $\delta \vec{v}$ describes separating trajectories, then
we can apply ray optics to describe the separation of the trajectories
\cite{Sinai1}. In the transverse plane this separation will be given
by an arc length equal to a
radius of curvature multiplied by an infinitesimal initial angular separation. We have chosen
the units in Eq. (\ref{28}) so that the radius of curvature operator has the
dimension of a length. The radius of curvature operator can be
represented as a $(d-1)\times(d-1)$ matrix since both $\delta\vec{r}_{\perp}$
and $\delta\vec{v}_{\perp}$ are defined in a plane perpendicular to
$\vec{v}$. Now suppose that we consider some initial
velocity deviation $\delta \vec{v}_{\perp}(0)$ corresponding to a diverging pencil of
rays, and we want to obtain an equation of
motion for the radius of curvature matrix $\mbox{\boldmath $\rho$}_{u}$. We
use the fact that the motion of the particle consists of periods of
free flight punctuated by instantaneous collisions with the fixed
scatterers. We first consider the free flight motion. From the fact
that in free flight $\delta\vec{r}_{\perp}(t)= \delta\vec{r}_{\perp}(0)+t \delta\vec{v}_{\perp}(0)$, and $\delta\vec{v}_{\perp}(t)=\delta\vec{v}_{\perp}(0)$, we infer that during free
flight over a time interval $t$ from some initial time $t=0$, with
initial value $\mbox{\boldmath $\rho$}(0)$, the radius of curvature matrix changes with time according to
\begin{equation}
\label{29}
 \mbox{\boldmath $\rho$}_{u}(t)=\mbox{\boldmath $\rho$}_u(0)+ vt{\bf 1}_{\perp}.
\end{equation}
where ${\bf 1}_{\perp}= {\bf 1}-\hat{v}\hat{v}$ with $\hat{v}$ a unit
vector in the direction of $\vec{v}$.
Next we use Eqs.(\ref{4a},\ref{4}) to obtain the
relation between the radius of curvature operator immediately before a
collision, $\mbox{\boldmath $\rho$}_{u}^{(-)}$, and its value immediately
after a collision, $\mbox{\boldmath $\rho$}_{u}^{(+)}$. This
calculation, while straight forward, requires a careful analysis in
order to obtain a correct expression for $\mbox{\boldmath$\rho$}$ as a
$(d-1)\times(d-1)$ matrix. This analysis is presented in Appendix
A. There we find that at a collision $\mbox{\boldmath$\rho$}$ changes
according to
\begin{equation}
[\mbox{\boldmath $\rho$}_{u}^{
-1
(+)}]
={\bf U} \left\{
[\mbox{\boldmath
$\rho$}_{u}^{
-1
(-)}]
+\frac{2}{a} \left[\hat{v}\hat{n}
+\hat{n}\hat{v}-\frac{1}{(\hat{v}\cdot\hat{n})}\hat{n}\hat{n}-(\hat{v}\cdot\hat{n}){\bf
1}
\right]
\right\}{\bf U}
,
\label{30}
\end{equation}
where ${\bf U}$ is the reflection operator ${\bf
1} -
2\hat{n}\hat{n}$. The inverse radius of curvature tensors
$[\mbox{\boldmath $\rho$}_{u}^{-1(-)}]$ and 
$[\mbox{\boldmath $\rho$}_{u}^{-1(+)}]$ are defined on the
subspaces orthogonal to $\hat{v}$ respectively ${\hat{v}}'$
and are extended to tensors on full space by requiring that
their left and right inner products with $\hat{v}$ 
respectively ${\hat{v}}'$ are zero again.

To obtain the connection between the radius of curvature operator and the
Lyapunov exponents for the Lorentz gas we proceed as follows. Suppose
we wish to know the 
values 
of the spatial separation of a bundle of
trajectories after a sequence of $n$ collisions labeled $1,2, ...,n$ that take place at
times $t_1, t_2, ...,t_n$. We use the fact that just before the
collision at time $t_n$, the spatial deviation vector is given by
\begin{equation}
\begin{array}{lcl}
\label{31}
\delta\vec{r}_{\perp}^{(-)}(t_n)&=&\delta\vec{r}_{\perp}^{(+)}(t_{n-1})+
\tau_{n,n-1}\delta\vec {v}_{\perp}^{(+)}(t_{n-1})\\
&=&\left[{\bf 1}_{\perp}+v\tau_{n,n-1}\mbox{\boldmath
$\rho$}_{u}^{
-1
(+)}(t_{n-1})
\right ]\cdot\delta\vec{r}_{\perp}^{(+)}(t_{n-1}),
\end{array}
\end{equation}
where $\tau_{i,i-1}=t_i-t_{i-1}$. We also use Eq.(\ref{4a}), namely 
that $\delta \vec{r}_{\perp}^{(+)} = {\bf U}\cdot
\delta \vec{r}_{\perp}^{(-)}$.
Then by iterating this equation we obtain
\begin{equation}
\begin{array}{lcl}
\label{32}
\delta\vec{r}_{\perp}^{(-)}(t_n) &=&\left[{\bf
1}_{\perp}+v\tau_{n,n-1}\mbox{\boldmath $\rho$}_{u}^{
-1
(+)}(t_{n-1})
\right]\cdot{\bf U}(n-1)\cdot  \\
& &\left[{\bf 1}_{\perp}+v\tau_{n-1,n-2}\mbox{\boldmath
$\rho$}_{u}^{
-1
(+)}(t_{n-2})
\right]\cdot{\bf U}(n-2)\cdots
\\
& &\left[{\bf 1}_{\perp}+v\tau_{1,0}\mbox{\boldmath
$\rho$}^
{-1}_u
(0)
\right]\cdot\delta\vec{r}_{\perp}(0),
\end{array}
\end{equation}
where the initial time has been set to
$t=0$, with initial values indicated for the spatial deviation vector,
$\delta\vec{r}_{\perp}(0)$ and for the radius of curvature operator $\mbox{\boldmath
$\rho
$}
_u
(0)$. Also ${\bf U}(j) = {\bf 1}-2\hat{n}_j\hat{n}_j$ is
the reflection operator at the $j$-th collision. It is important to
note that $\det {\bf U}(j)=-1$. 

If there is an exponential separation of trajectories,
then we would expect that $||\delta\vec{r}_{\perp}(t)|| \approx \left(\exp
\lambda t \right)||\delta\vec{r}_{\perp}(0)||$ for very large $t$. A
volume element constructed at time $t=0$ with  $d -1$ 
linearly independent
vectors  $\delta\vec{r}_{\perp}(0)^i$ is also expanding exponentially with
time. 
The exponential growth factor
is the sum of the positive Lyapunov
exponents and satisfies as consequence of  
Eq.\ ($\ref{32})$ 
\begin{equation}
\begin{array}{lcl}
\label{33}
\sum_{\lambda_i >0}\lambda_i &=& \lim_{t \rightarrow \infty}\frac{1}{t}\ln
\det \left[{\bf 1}_{\perp}+v\tau_{n,n-1}\mbox{\boldmath
$\rho$}_{u}^{
-1
(+)}(t_{n-1})
\right] \cdots \left[{\bf
1}_{\perp}+v\tau_{1,0}\mbox{\boldmath $\rho$}
_u^{-1}
(0)\right]
\\
&=&\lim_{t \rightarrow \infty} \frac{1}{t}\sum^{n-1}_{0}\ln \det \left[{\bf
1}_{\perp}+v\tau_{i+1,i}\mbox{\boldmath $\rho$}_{u}^{
-1
(+)}(t_i)
\right] \\
&=& \lim_{t \rightarrow
\infty}\frac{v}{t}\sum_{0}^{n-1}\int_{0}^{\tau_{i+1,i}}\,\, d \tau
{\rm Tr} \left[v
\tau
{\bf 1}_{\perp}+ \mbox{\boldmath
$\rho$}_{u}^{(+)}(t_i)\right]^{-1},
\end{array}
\end{equation}
where in the last expression the logarithm of a determinant is
expressed as an integral of a trace of 
an inverse matrix. With the aid of
Eq. (\ref{29}) we can express the sum of the positive Lyapunov
exponents as
\begin{equation}
\label{34}
\sum_{\lambda_i > 0}\lambda_i= \lim_{t \rightarrow \infty}
\frac{v}{t}\int_{0}^{t} \,\, d \tau {\rm Tr} \mbox{\boldmath
$\rho$}_{u}^{-1}(\tau).
\end{equation}

Eq. (\ref{34}) is Sinai's formula for the KS entropy for a moving
particle in a system of fixed hard-sphere scatterers\cite{Sinai1,Chernov}. By combining Eqs.\ (\ref{29}) and (\ref{30}) one may obtain a continued fraction
representation for $[\mbox{\boldmath $\rho$}^{-1}_{u}(t)]$ \cite{Sinai1}, which, for a fixed
final phase point $x(t)$ and initial $\mbox{\boldmath 
$\rho$}_{u}(0)$, converges
rapidly with increasing $t$. So far we have
not used any properties of the arrangement of the scatterers, so this
formula is still quite general. In the case that the system is
ergodic the time average can be replaced by an ensemble average,
taken with an appropriate ensemble distribution function, so
we can express the sum of the positive Lyapunov
exponents as
\begin{equation}
\sum_{\lambda_i > 0}\lambda_i= v\left<{\rm Tr} \mbox{\boldmath
$\rho$}_{u}^{-1}\right>,
\label{35}
\end{equation}
where the angular brackets denote an average over an appropriate
stationary ensemble reached in the course of time from 
smooth
initial distributions. In the case of interest here, this distribution
will be an equilibrium distribution but in 
subsequent papers
we will need to
consider more general steady state distribution functions. 

Before completing this section we wish to give a simple derivation of Eq. 
(\ref{34})
which applies to a Lorentz gas with any reasonable
interaction between the moving particle and the scatterers, and which
will be used often in the subsequent papers.  We use the fact that
having defined the radius of curvature matrix
$\mbox{\boldmath$\rho$}_u$, we may write
\begin{eqnarray}
\frac{d \delta \vec{r}_{\perp}(t)}{dt} = \delta \vec{v}_{\perp}(t)
\nonumber  \\
= v [\mbox{\boldmath$\rho$}_{u}
^{-1}
(t)]
\cdot \delta
\vec{r}_{\perp}(t),
\label{35a}
\end{eqnarray}
with solution
\begin{equation}
\delta \vec{r}_{\perp}(t) = {\cal T}\exp v{\int_0^{t} d \tau
[\mbox{\boldmath$\rho$}_{u}(\tau)]^{-1}}\cdot \delta \vec{r}_{\perp}(0).
\label{35b}
\end{equation}
Here ${\cal T}$ denotes a time-ordering operator.
Using the method of differential forms or equivalent methods \cite{Flanders} we see
that the growth of a volume element, $\delta V_{r}(t)$, in configuration space is given by
\begin{equation}
\frac{\delta V_{r}(t)}{\delta V_{r}(0)} = \exp \left( v\int_0^t d \tau {\rm
Tr} [\mbox{\boldmath$\rho$}_{u}(\tau)]^{-1}\right).
\label{35c}
\end{equation}
This result leads immediately to Eq.(\ref{34}) for the sum of Lyapunov
exponents. Vattay has shown how to construct the inverse of the radius
of curvature matrix for a general potential \cite{Vattay}.
It is straightforward generalizing
the results obtained here to other
short range 
interaction
potentials
and it may well be possible to treat some cases with long range interactions between the scatterers and the light particle. 
For the case of hard disks or spheres considered here, we
can use the fact that the radius of curvature matrices at different
times commute with
each other if the times involved are all within the same time interval
between one collision of the moving particle and a scatterer and
the next collision, to write
\begin{equation}
\delta \vec{r}_{\perp}(t) = \left[{\prod_{j=0}^{n}}^{\prime} {\bf
R}(t_{j+1},t_j)\right]\cdot \delta \vec{r}_{\perp}(0),
\label{35d}
\end{equation}
where
\begin{equation}
{\bf R}(t_{j+1},t_j) = \exp \left[v\int_{t_j}^{t_{j+1}}d \tau
\left[\mbox{\boldmath$\rho$}
_u(\tau)\right]^{-1}\right],
\label{35e}
\end{equation}
and the prime on the product denotes that the times are to be ordered
so that the times decrease from left to right in the product. By using
Eq. (\ref{29}) and carrying out the required integrals, one can easily see
that this expression is equivalent to Eq. (\ref{32}). Moreover one can
express the sum of the positive Lyapunov exponents as
\begin{equation}
\sum\lambda_{i}^+ =
\lim_{t\rightarrow\infty}\frac{{\cal N}(t)}{t}\frac{1}{{\cal N}(t)}\sum_{j}v\int_{t_j}^{t_{j+1}}d\tau
{\rm Tr}\left[\mbox{\boldmath$\rho$}
_u(\tau)\right]^{-1}
\label{35f}
\end{equation}
We will use these expressions in Section VI and in the appendices.

 In the next
section we discuss the  distribution function appearing in the above
ensemble average for the case that the
scatterers are distributed at random with very low density, i.e., the mean
free path of the moving particle is very large compared to the radius
of the scatterers. In order to obtain the individual Lyapunov
exponents we have to find the eigenvalues of the operator that appears
on the right hand side of Eq. (\ref{32}). This operator can be
expressed as a product of $(d-1)\times (d-1)$ matrices, which describe the
collisions of the moving particle with the scatterers, and the free
motion in between collisions. Again, if the system is at low density
the product of matrices can be considered to be a product of randomly
distributed matrices, since the time between collisions and the
collision parameters will be sampled from a random distribution,
corresponding to the random placement of scatterers. In Appendix B we
show how methods from the theory for eigenvalues of products of random matrices
\cite{Vulp} can be used to obtain the largest Lyapunov exponent.  At higher
densities of scatterers
correlations between collisions will arise,
and
one will have to take into account 
these correlations 
when
computing the  eigenvalues of the product of matrices.

\section{The extended Lorentz-Boltzmann Equation}

In order to evaluate the ensemble average appearing on the right hand
side of Eq. (\ref{35}) we need to construct an equilibrium, or more
generally, a steady state distribution function for
the radius of curvature matrix. The physics of the problem suggests
that a method based upon the Lorentz-Boltzmann equation is appropriate
here. That is, we have a particle moving in a random array of
scatterers making only binary collisions with the scatterers. The
particle moves freely between collisions and at a
collision both the velocity and the radius of curvature matrix change
instantaneously. Methods are now well known \cite{Hauge,Cohen1,Chap-Cow,Do-HVB,hvl-weij} for obtaining a
generalized Lorentz-Boltzmann equation for the time dependent space
and velocity distribution function, $f(\vec{r},\vec{v},t)$, for the moving particle as a
function of the density of the scatterers, at least in the case that
the scatterers are non-overlapping \footnote{The case of overlapping
scatterers is complicated by the fact that regions may exist where the
particle would be trapped for all time. In a transport problem these
regions need to be treated carefully, since particles trapped in them
will not diffuse beyond the borders of the trap.}. In this case it is
possible to obtain an equation for the moving particle which includes
the effects of uncorrelated binary collisions of the particle with the
scatterers, excluded volume effects and the effects of correlated
collision sequences of the moving particle with the scatterers. To
lowest order in the density of the scatterers, the distribution
function $f(\vec{r},\vec{v},t)$, satisfies the Lorentz-Boltzmann (LB)
equation\cite{Hauge,Cohen1,Chap-Cow,hvl-weij}
\begin{eqnarray}
&& \frac{\partial f}{\partial t} +\vec{v}\cdot\frac{\partial f}{\partial
\vec{r}} + \dot{\vec{v}}\cdot \frac{\partial f}{\partial \vec{v}}
= \nonumber \\
&& na^{d-1}\int
d\,\hat{n}\,|\vec{v}\cdot\hat{n}|
[\Theta(\vec{v}\cdot\hat{n})f(\vec{r},\vec{v}-2
(\vec{v}\cdot\hat{n})\hat{n},t)-\Theta(-\vec{v}\cdot\hat{n})f(\vec{r},\vec{v},t)].
\label{36}
\end{eqnarray}
Here $\hat{n}$ is a unit vector in the direction from the center of a scatterer
to the point of impact at a collision, and 
$\Theta(x)$ denotes the unit step function.
The right hand side of the LB
equation describes the change in $f$ due to collisions as the
difference between the gain and the
loss of particles with velocity $\vec{v}$ from ``restituting" and
``direct" collisions, respectively. Higher order density corrections to the
right hand side of Eq. (\ref{36}), comprising the generalized LB
equation, can be obtained using the appropriate set of BBGKY hierarchy
equations and cluster expansion methods\cite{Do-HVB,hvl-weij,Cohen2}. These density corrections
have been studied in some detail, and it is well known that the
so-called ring collision sequences are responsible for both
logarithmic terms in the density expansion of the diffusion
coefficient of the moving particle\cite{hvl-weij} 
and
the long time tails in
the velocity auto-correlation function
of the moving particle\cite{ern-weij}.

Our purpose here is to extend the LB equation by including the
radius of curvature matrix among the variables described by the
distribution function for the moving particle. We will consider only
the low density version of the kinetic theory for this distribution
function and leave the discussion of higher order density corrections,
including the effects of correlated collision sequences on the Lyapunov exponents,
to later publications. Thus we wish to determine an equation,
valid at low densities,
for an
extended distribution function $F(\vec{r},\vec{v},\mbox{\boldmath
$\rho$},t)$, 
where we dropped the subscript $u$ on $\mbox{\boldmath
$\rho$}$, 
and we relate the distribution
functions $F$ and $f$ by
\begin{equation}
f(\vec{r},\vec{v},t) = \int d\,
\mbox{\boldmath$\rho$}\,F(\vec{r},\vec{v},\mbox{\boldmath$\rho$},t).
\label{37}
\end{equation}
Given the stationary solution of the extended LB equation for $F$, we can
determine the sum of the Lyapunov exponents as
\begin{equation}
\sum_{\lambda_i >0}\lambda_i =v\int
d\vec{r}\,d\,\vec{v}\,d\,\mbox{\boldmath$\rho$}\left[{\rm
Tr}\mbox{\boldmath$\rho$}^{-1}\right]\,F(\vec{r},\vec{v},\mbox{\boldmath$\rho$}),
\label{38}
\end{equation}
assuming that $F$ is properly normalized.

An extended LB equation for $F$ that reduces to the usual LB equation
for $f$ upon integration over the radius of curvature matrix elements,
can be obtained by following the heuristic derivation of the LB
equation and simply modifying it to include the additional
variables. That is, we consider a large collection of moving particles
in the random array of scatterers and ask for an equation for the probability that a
moving particle has its values for $\vec{r}, \vec{v},
\mbox{\boldmath$\rho$}$ in the range
$d\vec{r},d\vec{v},d\mbox{\boldmath$\rho$}$ about $\vec{r},\vec{v},
\mbox{\boldmath$\rho$}$, all at time $t$, i.e., $F(\vec{r}, \vec{v},
\mbox{\boldmath$\rho$},t)d\vec{r}d\vec{v}d\mbox{\boldmath$\rho$}$.
This probability changes in time due to free motion of the particles
and due to collisions. The change in $F$ due to the free motion of the
particles in time $dt$ is
\begin{equation}
\begin{array}{lcl}
[F(\vec{r}+\vec{v}dt, \vec{v}, \mbox{\boldmath$\rho$}+vdt{\bf 1}_{\perp},t+dt)-F(\vec{r}, \vec{v},
\mbox{\boldmath$\rho$},t)]d\vec{r}\,d\vec{v}\,d\mbox{\boldmath$\rho$}&=& \\
\left[\frac{\partial
F}{\partial t} + \vec{v}\cdot\frac{\partial F}{\partial \vec{r}}
+v\sum_{i=1}^{d-1}\frac{\partial F}{\partial
\rho_{ii}}\right] d\vec{r}\,d\vec{v}\,d\mbox{\boldmath$\rho$}\,dt.& &
\label{39}
\end{array}
\end{equation}
We used Eq. (\ref{29}) to treat the change in the radius of curvature matrix
during free particle motion, and we have assumed that there are no external forces acting on the
system. Otherwise we would need to include terms accounting for the
changes in velocity and in the radius of curvature matrix over a time interval $dt$ due to the external
force. 
If there were no collisions taking place in the system, then the right
hand side of Eq. (\ref{39}) would be zero. However the collisions
account for the fact that the number of particles at
$\vec{r}+\vec{v}dt, \vec{v}, \mbox{\boldmath$\rho$}+vdt{\bf 1}_{\perp}$ at
time $t+dt$ is not equal to the number of particles at $\vec{r}, \vec{v},
\mbox{\boldmath$\rho$}$ at time $t$. To account for the change in $F$
due to collisions we consider the restituting and direct collisions
separately. The direct collisions result in a loss of the particles
with $\vec{r}, \vec{v},
\mbox{\boldmath$\rho$}$ over the interval $dt$ due to collisions with
scatterers. Elementary kinetic theory considerations \cite{Chap-Cow,Do-HVB} show that this
loss is
\begin{equation}
na^{d-1}\int d\hat{n}
\left| \vec{v}\cdot\hat{n}\right|\Theta(-\vec{v}\cdot\hat{n})
F(\vec{r}, \vec{v},
\mbox{\boldmath$\rho$},t)
 d \vec{r} \, d \vec{v} \,d \mbox{\boldmath$\rho$}\,dt 
.
\label{40}
\end{equation}
The restituting, or gain
term is
found by considering those collisions taking place in the time interval
$t$ to $t+dt$ that produce particles with $\vec{r}, \vec{v},
\mbox{\boldmath$\rho$}$. Again, elementary kinetic theory
considerations show that this gain is given by
\begin{equation}
na^{d-1}\int d\hat{n}
\left|\vec{v}\cdot\hat{n}\right|\Theta(\vec{v}\cdot\hat{n})\,\left|\frac{\partial
\mbox{\boldmath$\rho$}'}{\partial
\mbox{\boldmath$\rho$}}\right|\,F(\vec{v}',\vec{r},\mbox{\boldmath$\rho$}',t)d\vec{r}\,d\vec{v}\,d\mbox{\boldmath$\rho$}\,dt
,
\label{41}
\end{equation}
where $\vec{v}\cdot\hat{n} \geq 0$, and
\begin{equation}
\vec{v}'=\vec{v}-2(\vec{v}\cdot\hat{n})\hat{n},
\end{equation}
 and the radius of curvature matrix $\mbox{\boldmath$\rho$}'$ is a
restituting value such that the radius of curvature matrix becomes
$\mbox{\boldmath$\rho$}$ after collision.
From Eq.\ (\ref{30}) and the identity ${\bf U}\cdot\hat{v}=
{\hat{v}}'$ one obtains the relationship
\begin{equation}
{\mbox{\boldmath$\rho$}}'^{-1}=
{\bf U}\left\{\mbox{\boldmath$\rho$}^{-1} -\frac{2}{a}
\left[-\hat{v}\hat{n}-\hat{n}\hat{v}+\frac{1}{(\hat{v}\cdot\hat{n})}
\hat{n}\hat{n}+(\hat{v}\cdot\hat{n}){\bf
1}
\right]
\right\}{\bf U}
\label{42}
\end{equation}
with $\hat{v}\cdot\hat{n}\ge 0$.
For two dimensional systems the radius of curvature matrix can be
represented by a
scalar (namely its non-vanishing eigenvalue), and the restituting value, 
$\rho'$,
is given by
\begin{equation}
\frac{1}{\rho'}=\frac{1}{\rho}-\frac{2}{a\cos\phi}
\label{43}
\end{equation}
where $\phi$ is the angle between $\hat{n}$ and $\hat{v}$ with $-\pi/2
\leq \phi \leq \pi/2$. For three dimensional systems the radius of
curvature matrix 
can be represented by
a $2\times2$ matrix by choosing the principal axes of the
coordinate frame orthogonal to ${\hat{v}}$. Defining angles
$\phi$ through $\cos\phi = \hat{n}\cdot\hat{v}$ with $0\leq 
\phi\leq \pi/2$ and $\alpha$ as the angle between the second 
coordinate axis and the plane through $\hat{n}$ and
$\hat{v}$, and multiplying Eq.\ (\ref{42}) from the left and 
the right by ${\bf U}$ one can rewrite this equation as
\begin{equation}\label{44}
{\bf U}
{\mbox{\boldmath$\rho$}'}^{-1}
{\bf U}
={\mbox{\boldmath$\rho$}}^{-1}
-\frac{2\cos\phi}{a} \left(\begin{array}{cc}
0&\begin{array}[t]{cc}
0&0
\end{array}\\
\begin{array}{l}
0\\
0
\end{array}&{\bf P}
\end{array}\right)
\end{equation}
with
\begin{equation}\label{defp}
{\bf P} = \left(\begin{array}{cc}
1
+\tan^{2}\phi\cos^{2}\alpha &  \tan^{2}\phi\sin\alpha\cos\alpha \\
\tan^{2}\phi\sin\alpha\cos\alpha & 
1
+\tan^{2}\phi\sin^{2}\alpha
\end{array}\right).
\end{equation}
Here $\cos\phi = \hat{n}\cdot\hat{v}$ with $0\leq \phi\leq \pi/2$ and
$\alpha$ is an azimuthal angle for $\hat{n}$ in the plane
perpendicular to $\vec{v}$ with $0\leq \alpha \leq 2\pi$. We can now
simplify the restituting term in the extended LB equation by noting
that we can combine the Jacobian with the distribution function $F$ to
write
\begin{equation}
\left| \frac{\partial \mbox{\boldmath$\rho$}'}{\partial
\mbox{\boldmath$\rho$}} \right| F(\mbox{\boldmath$\rho$}')=\int\,
d\,\mbox{\boldmath$\rho$}\prime \delta\left(
\mbox{\boldmath$\rho$}-\mbox{\boldmath$\rho$}(\mbox{\boldmath$\rho$}\prime)\right)\,F(\mbox{\boldmath$\rho$}\prime),
\label{45}
\end{equation}  
where the $\delta$ function in the integrand selects the right
restituting value of the radius of curvature matrix in accordance with
Eq. (\ref{42}). Putting everything together, we can obtain the
extended LB equation as
\begin{eqnarray}
 &&\frac{\partial F}{\partial t}+\vec{v}\cdot\frac{\partial
F}{\partial \vec{r}}+v\sum_{i=1}^{d-1}\frac{\partial F}{\partial
\rho_{ii}} = \nonumber \\
&&na^{d-1}\int
d\hat{n}|\vec{v}\cdot\hat{n}|\left[
\Theta(\vec{v}\cdot\hat{n})
\int\,d\,\mbox{\boldmath$\rho$}'
\delta\left(\mbox{\boldmath$\rho$}-\mbox{\boldmath$\rho$}(\mbox{\boldmath$\rho$}')\right)F(\vec{r},\vec{v}',\mbox{\boldmath$\rho$}',t)
- 
\Theta(-\vec{v}\cdot\hat{n})
F(\vec{r},\vec{v},\mbox{\boldmath$\rho$},t)\right].
\label{46}
\end{eqnarray}
In the next section we
will use Eq. (\ref{46}) to compute the KS entropy for two and three
dimensional Lorentz gases in equilibrium. Before turning to this
calculation, we make an observation about the restituting radius of
curvature matrix. We note that the diagonal
elements of the radius of curvature matrix will grow with time between
collisions. Thus the average value of the diagonal elements of the
radius of curvature matrix {\em before collision} will be on the order
of the mean free path between collisions, $l$. For low density of
scatterers, the mean free path, $l$, will be much larger than the radius of
the individual scatterers, $a$, such that $a/l \sim na^d \ll 1$. This
observation will allow us to greatly simplify the delta function in the
restituting collision term in Eq. (\ref{46}), and thereby simplify the
calculations to follow.

\section{Equilibrium Solutions of the Extended LB Equation}

In this section we construct the equilibrium solutions of the extended
LB equation, Eq. (\ref{46}), in two and three dimensions and from
these compute $h_{KS}$. We begin with the two dimensional case. Here
the radius of curvature is a simple scalar, and Eq. (\ref{46}) becomes
\begin{eqnarray}
&&\frac{\partial F}{\partial t} +\vec{v}\cdot\frac{\partial F}{\partial
\vec{r}}+v\frac{\partial F}{\partial \rho}= \nonumber \\
&&nav\int_{-\pi/2}^{\pi/2}d\phi \cos\phi\left[\int_{0}^{\infty}d\rho'
\delta\left(\rho-\frac{a\cos\phi}{2
+\frac{a\cos\phi}{\rho'}}\right)F(\vec{r},
\vec{v}',\rho',t)-F(\vec{r},\vec{v},\rho,t)\right].
\label{47}
\end{eqnarray}
To find an equilibrium solution, we look for solutions that do not
depend upon time, velocity direction, or position, and which become the known equilibrium
solution for the LB equation when the integration over $\rho$ is
carried out. That is we look for solutions $F$ of the form
\begin{equation}
F(\vec{v},\rho)=\varphi_0(v)\psi(\rho)
\label{48}
\end{equation}
where $\varphi_0(v)=(2\pi v V)^{-1}\delta(v-v_0)$ is the normalized, equilibrium
spatial and velocity distribution function for the moving particle
with constant speed, which we here denote as $v_0$, and confined to a
volume $V$. Then we require
that $\psi(\rho)$ be normalized as
\begin{equation}
\int_0^{\infty}d\rho \psi(\rho) =1.
\label{49}
\end{equation}
It is an easy matter to obtain an equation for $\psi(\rho)$ which
reads
\begin{equation}
v\frac{\partial \psi}{\partial \rho}=-2nav\psi
+nav\int_{-\pi/2}^{\pi/2}d\phi\int_0^{\infty}d\rho'\delta\left(\rho-\frac{a\cos\phi}{2+\frac{a\cos\phi}{\rho'}}\right)\psi(\rho').
\label{50}
\end{equation}
An inspection of the delta function shows that it vanishes unless
$\rho < a/2$. Therefore, for $\rho \geq a/2$, we have the simple
result
\begin{equation}
\psi(\rho)=Ae^{-\frac{\rho}{\ell}} \,\,\, {\rm for}
\ 
\rho\geq a/2
\label{51}
\end{equation}
where $\ell = (2nav)^{-1}$ is the mean free path length for the moving
particle at low density of scatterers, and $A$ is a constant to be determined. To treat the distribution
function $\psi(\rho)$ for smaller values of $\rho$ we note that we can
require that $\psi(\rho)\rightarrow 0$ as $\rho\rightarrow 0$, since
the dynamics will increase the value of $\rho$ during free particle
motion and will decrease it to some value, still greater than $0$, at the instant
of a collision. Further we can require that $\rho$ be continuous at
$\rho =a/2$ since the extended LB equation does not have an
explicit delta function of the form $\delta(\rho - a/2)$ on the right
hand side. Finally we note that the dominant contribution to the
$\rho'$ integral on the right hand side of Eq. (\ref{50}) comes from
$
\rho' \sim \ell$. Therefore the delta function on the right hand side
of Eq. (\ref{50}) can be approximated by\footnote{A more detailed
examination of this integral keeping the full delta function shows
that the terms neglected here are of order $\tilde{n}\ln\tilde{n}$
compared to the terms retained, where $\tilde{n}=na^2$.}
\begin{equation}
\delta\left(\rho -\frac{a\cos\phi}{2
+\frac{acos\phi}{\rho'}}\right)\simeq \delta(\rho -(a\cos\phi)/2).
\label{52}
\end{equation} 
After inserting this expression in Eq. (\ref{50}) we find that for
$\rho < a/2$,
\begin{equation}
v\frac{\partial \psi}{\partial \rho} = 2nav\left[- \psi(\rho)
+\frac{2\sigma}{a(1-\sigma^2)^{1/2}}\int_0^{\infty}d\rho'\psi(\rho')\right],
\label{53}
\end{equation}
where $\sigma = a\rho/2$. Since $\psi$ is normalized to one, we find 
that\footnote{Comparing Eqs.~(\ref{51}) and (\ref{54}) one sees that 
$\psi(\rho)$ apparently has a discontinuity at $\rho=a/2$. This, however, is
an artifact of the low density approximations we have made. Note that the jump
in $\psi$ is of relative order $na^2$ indeed.}
\begin{equation}
\psi(\rho) = (1/\ell)\left[ 1 - \left( 1 -
\left(\frac{2\rho}{a}\right)^2\right)^{1/2}\right]\,\,\, {\rm for}\,\,
\rho < a/2,
\label{54}
\end{equation}
and, using the normalization condition on $\psi$, we find
\begin{equation}
\psi(\rho) = (1/\ell)e^{-\frac{\rho}{\ell}}\,\,\, {\rm for}\,\,
\rho\geq a/2.
\label{55}
\end{equation}
Combining this expression for $\psi$ with Eqs. (\ref{38},
\ref{48}) we obtain
\begin{equation}
\lambda^{(+)}=h_{KS}=2nav(1-\ln 2 -{\cal C}-\ln \tilde{n}) \,\,\,{\rm
for}\,\,\tilde{n} \ll 1,
\label{56}
\end{equation}
in agreement with the result given by Eq. (\ref{22}).

Now we turn to the three dimensional case. This is somewhat more
complicated than the two dimensional case since
$\mbox{\boldmath$\rho$}$ is a $2\times2$ matrix and not a
scalar. However, we can still simplify the delta function in the restituting
collision integral by noticing that the diagonal elements of the
curvature matrix grow with time during the free steaming intervals
between collisions. Consequently, the diagonal elements of
$\mbox{\boldmath$\rho$}^\prime$ appearing on the left hand side of Eq.
(\ref{44}) are of the order of the mean free path length
immediately before a collision with a scatterer. An elementary
consideration of the properties of the inverses of $2\times 2$
matrices with large diagonal elements shows that the dominant
contribution to the radius of curvature matrix $\mbox{\boldmath
$\rho$}$ comes from setting the left hand side of Eq. (\ref{44}) equal
to zero. This greatly simplifies the delta function appearing in the
collision integral on the right hand side of Eq. (\ref{46}). The
effect of this simplification is that we have neglected terms of
relative order $a/\ell$, which are density corrections to the terms we
keep. 

The equilibrium distribution
function $F(\vec{v},\mbox{\boldmath$\rho$})$ 
can be factorized as
\begin{equation}
F(\vec{v},\mbox{\boldmath$\rho$})=\varphi(v)\psi(\mbox{\boldmath$\rho$}),
\end{equation}
where $\varphi(v)=(4\pi^{2}v_{0}^{2}V)^{-1}\delta(v-v_0)$ is the
normalized equilibrium distribution function for the moving particle.
The extended LB equation 
reduces to
\begin{equation}
\begin{array}{l}
v\left(\frac{\partial}{\partial \rho_{11}} +\frac{\partial}{\partial
\rho_{22}}\right)\psi(\mbox{\boldmath$\rho$}) = -na^{2}\pi v
\psi(\mbox{\boldmath$\rho$}) + \\
+na^{2}v\int_{0}^{\pi/2}d\phi\int_{0}^{2\pi}d\alpha\int d
\rho'_{11}\int d\rho'_{12}\int d\rho'_{21}\int
d\rho'_{22}\sin\phi\cos\phi
\prod_{i,j}\delta(\rho_{ij}-\rho_{ij}(\phi,\alpha))
\, \psi(\mbox{\boldmath$\rho$}').
\label{57}
\end{array}
\end{equation}
The matrix elements of $\mbox{\boldmath $\rho
(\phi,\alpha)$}$ can be obtained by
solving Eq. (\ref{44}) under the approximation 
$[\mbox{\boldmath $\rho$}']^{-1}=0$, which can be justified again 
as a low density approximation by the same arguments as in 
the two dimensional case. 
The results are
\begin{eqnarray}
\rho_{11}(\phi,\alpha)=\frac{a\cos\phi}{2
}(
1+\tan^2\phi\sin^2\alpha)
\\
\rho_{12}(\phi,\alpha)=\rho_{21}(\phi,\alpha)=
-\frac{a\sin^2\phi\sin2\alpha}{4
\cos\phi}
\label{rho12}
\\
\rho_{22}(\phi,\alpha)=\frac{a\cos\phi}{2
}(
1+\tan^2\phi\cos^2\alpha)
\label{58}.
\end{eqnarray}
The
restituting term in Eq. ({\ref{57}) contains an integration of
$\psi(\mbox{\boldmath$\rho$}')$ over its 
arguments, and we can set
\begin{equation}
\int d \mbox{\boldmath$\rho$}' \psi(\mbox{\boldmath$\rho$}') = 1
\label{59}
\end{equation} 

We use $\psi$ to compute the average value of ${\rm
Tr}(\mbox{\boldmath$\rho$})^{-1}=[{\rm
Tr}\mbox{\boldmath$\rho$}][\det\mbox{\boldmath$\rho$}]^{-1}$, which
determines the sum of the positive Lyapunov exponents. 
The $\rho_{ij}(\phi,\alpha)$ occurring in the delta function
on the right hand side of Eq.\ (\ref{57}) can be identified
at low density with the values of $\rho_{ij}$ right after
a collision with collision parameters $\phi$ and $\alpha$.
From Eq. (\ref{rho12}) it follows that these values are 
always equal for $\rho_{12}$ and $\rho_{21}$. Furthermore
these quantities do not change in between collisions, so we
may set 
$\psi(\mbox{\boldmath$\rho$})=f(\rho_{11},\rho_{22},\rho_{12})\delta(\rho_{12}-\rho_{21})$.
Next we change variables from $\rho_{11},\rho_{22},\rho_{12}$ to
$\rho_1,\rho_2,\rho_{12}$ where $\rho_1,\rho_2$ are the eigenvalues of
the $2\times2$ matrix $\mbox{\boldmath$\rho$}$. The Jacobian for the
transformation of variables, given by 
\begin{equation}
J\left(\frac{\rho_1,\rho_2}{\rho_{11},\rho_{22}}\right) =
\frac{|\rho_1-\rho_{2}|}{[(\rho_{1}-\rho_{2})^{2}-4\rho_{12}^{2}]^{1/2}},
\label{59a}
\end{equation}
 will be included in the integrand in Eq. (\ref{62}). After some
straightforward algebra
we obtain the following equation for
$g(\rho_1,\rho_2,\rho_{12}) \equiv f(\rho_{11},\rho_{22},\rho_{12})$:
\begin{equation}
\begin{array}{l}
v\left(\frac{\partial}{\partial \rho_1}+\frac{\partial}{\partial
\rho_2}\right)g(\rho_1,\rho_2,\rho_{12}) +\nu
g(\rho_1,\rho_2,\rho_{12}) = \\
\frac{\nu}{\pi}\int_{0}^{\pi/2} d\phi \int_{0}^{2\pi} d\alpha \sin \phi
\cos \phi |\cos
2\alpha | \delta \left(\rho_1-\frac{a}{2\cos\phi}\right) \delta\left(\rho_2
-(a\cos\phi)/2\right)\\
\delta\left(\rho_{12}+(a\cos\phi\tan^{2}\phi\sin2\alpha)/4\right)
\label{60}.
\end{array}
\end{equation}
where $\nu =n\pi a^{2}v$ is the average collision frequency for the
moving particle. The $\alpha$ integration can be carried out and we see that $g$ can be
written in the form
\begin{equation}
g(\rho_1,\rho_2,\rho_{12}) =\Theta\left(1-\frac{2|\rho_{12}|}{|\rho_1
-\rho_2|}\right) h(\rho_1,\rho_2)
\end{equation}
where 
$h$ satisfies
\begin{equation}
\begin{array}{l}
v\left(\frac{\partial}{\partial \rho_1}+\frac{\partial}{\partial
\rho_2}\right)h(\rho_1,\rho_2)+\nu
h(\rho_1,\rho_2)=\\
\frac{
4\nu}{a\pi}\int_0^{\pi/2}d\phi
\frac{\cos^2\phi}{\sin\phi}\delta(\rho_
2-(a\cos\phi)/2)\delta(\rho_
1-a(2\cos\phi)^{-1}).
\label{61}
\end{array}
\end{equation}
We can now express the sum of the positive Lyapunov exponents in terms
of $h$ as
\begin{equation}
\begin{array}{l}
\lambda_{max}^{+}+\lambda_{min}^{+}=\\
\int d\rho_1 \int d\rho_2 \int
d\rho_{12}\left(\frac{1}{\rho_1}
+\frac{1}{\rho_2}\right)\left(\frac{|\rho_1-\rho_2|}{[(\rho_2 -\rho_1)^{2}-4\rho_{12}^{2}]^{1/2}}\right)\Theta\left(1-\frac{2|\rho_{12}|}{|\rho_1
-\rho_2|}\right) h(\rho_1,\rho_2).
\label{62}
\end{array}
\end{equation}
The $\rho_{12}$ integration can be carried out easily yielding
\begin{equation}
\lambda_{max}^{+}+\lambda_{min}^{+}=\frac{\pi}{2}\int d\rho_1 \int d\rho_2 \left(\frac{1}{\rho_1}
+\frac{1}{\rho_2}\right)|\rho_2 -\rho_1|h(\rho_1,\rho_2).
\label{63}
\end{equation}
We define a new function $p(\rho_1,\rho_2)=(\pi/2)|\rho_2 -
\rho_1|h(\rho_{1}, \rho_2)$ which satisfies
\begin{equation}
\begin{array}{l}
v\left(\frac{\partial}{\partial \rho_1}+\frac{\partial}{\partial
\rho_2}\right)p(\rho_1,\rho_2) + \nu p(\rho_1,\rho_2)=\\
\nu\int_{0}^{\pi/2}d\phi
\frac{\cos^2\phi}{\sin\phi}\left(\frac{1}{\cos\phi}-\cos\phi\right)\delta(\rho_
2-(a\cos\phi)/2)\delta(\rho_
1-a(2\cos\phi)^{-1}).
\label{
63a}
\end{array}
\end{equation}
Introducing $p_1(\rho)=\int_0^\infty d\rho_2 
p(\rho,\rho_2)$ and $p_2(\rho)=\int_0^\infty d\rho_1 
p(\rho_1,\rho)$ enables us to express the sum of the Lyapunov exponents very
simply as
\begin{equation}
\lambda_{max}^{+}+\lambda_{min}^{+}=\sum_{i=1,2}\int_{0}^{\infty}d\rho
\frac{1}{\rho}\,\,p_{i}(\rho)
\label{64}
\end{equation}
where the $ p_{i}(\rho) $ satisfy
\begin{eqnarray}
v\frac{\partial p_{1}(\rho)}{\partial \rho} +\nu
p_{1}(\rho)&=&2\nu\int_{0}^{\pi/2}d\phi
\sin\phi\cos\phi\delta(\rho-(a\cos\phi)/2) \,\,\,{\rm and} \label{65a}\\
v\frac{\partial p_{2}(\rho)}{\partial \rho} +\nu
p_{2}(\rho)&=& 2\nu\int_{0}^{\pi/2}d\phi
\sin\phi\cos\phi\delta(\rho- a(2\cos\phi)^{-1}).
\label{65}
\end{eqnarray}
Solving Eqs. (\ref{65a}, \ref{65}) 
for $p_1, p_2$ and
inserting the 
solution 
into
Eq. (\ref{64})
one obtains an expression for the sum of the positive Lyapunov exponents
which agrees with that 
of Eq. (\ref{23}). It is worth
mentioning that we can solve Eq. (\ref{60}) to provide $g(\rho_1,\rho_2,
\rho_{12})$ as
an explicit function of the variables $\rho_1,\rho_2, \rho_{12}$ using 
the method of characteristics. As this
solution will prove useful in
subsequent papers, we outline the
method in Appendix B. There we also briefly indicate how the
individual Lyapunov exponents can be calculated using $g(\rho_1,
\rho_2, \rho_{12})$ and simple results from the theory for eigenvalues of
products of random matrices \cite{Vulp}. While the results obtained with the
extended LB equation properly agree with those obtained by more direct
kinetic theory arguments, we will need to use the extended LB equation
in order to obtain the Lyapunov exponents and KS entropies for the
spatially inhomogeneous systems that occur when one considers
escape-rate methods for connecting chaotic quantities with transport
coefficients. This will be treated in a subsequent paper.

\section{The Negative Lyapunov Exponents and the Anti-Lorentz
Boltzmann Equation.}

We now turn our attention to the Lyapunov exponents that characterize
the exponential convergence of trajectories on a stable manifold in
the $2d-1$ dimensional constant energy surface in the phase space of
the moving particle. We recall that two arbitrary but
infinitesimally nearby trajectories will certainly separate eventually
with time. We have used this fact to derive formulas and explicit
expressions for the positive Lyapunov exponents. However Liouville's
theorem showing that the measure of a small region of phase space
is constant in time as one follows the motion of points initially in
that region implies that there must be a compensating set of negative
Lyapunov exponents which act in concert with the positive ones to keep
phase space measures constant in time. Moreover, the fact that the
Lorentz gas is a symplectic Hamiltonian system
has as a consequence
the existence of a conjugate pairing rule\cite{Arnold}. That is for such a system,
the Lyapunov exponents come in positive and negative pairs such that
the sum of each corresponding pair
is zero. Thus, in
this case at least, the calculation of the negative Lyapunov exponents
is trivial - they are just the 
opposites of the positive ones. However,
in a subsequent paper where we treat thermostatted systems, this form
of the conjugate pairing rule no longer holds\cite{EvMoCo,DettMo,Woj-Liv} and we will need to find
methods to compute
both positive and
negative exponents individually.

The most obvious way to obtain the negative Lyapunov exponents is
to compute the positive Lyapunov exponents {\em for the time reversed
motion}. 
Upon time reversal
trajectories that approach each
other in the forward motion will separate.
In fact almost all trajectories will separate in {\em both} the forward and
the
time reversed 
direction, but in general (i.e. for non-symplectic
systems) with different exponents.
In the forward motion 
they will separate 
with rates given by the positive Lyapunov exponents, and in the time reversed motion
with rates 
equal to the magnitudes of the negative Lyapunov
exponents. Thus to calculate the negative exponents we consider the
binary collision dynamics already discussed 
before, but look at the time
reversed motion. If,
in the forward time direction, the moving particle
is uncorrelated with a scatterer 
before
collision,
in the time reversed motion 
it will be uncorrelated with
the same scatterer after the collision.
We 
therefore 
should consider a kind of backwards kinetic theory,
where
the particles are uncorrelated with the scatterers after their
collisions instead of before them. 
This will
differ in important respects from the ordinary Lorentz-Boltzmann equation.
In order to illustrate 
this we
consider first a calculation of the sum of the negative Lyapunov exponents for the
2-d and 3-d dilute Lorentz gases, using a simple kinetic theory
argument similar to that used in Section II.

\subsection{ A Simple Kinetic Theory Method for the Sum of the
Negative Lyapunov Exponents} 

We first consider the 2-d case. We wish to follow an infinitesimal trajectory bundle
which is contracting and remains contracting for all future times.  
Such a bundle is illustrated in Fig. 1. We follow the motion of this
bundle from scatterer $1$ to scatterer $2$. In order that the bundle
remain contracting after the collision with scatterer $2$ we require
that the radius of curvature of this bundle be very close (with
corrections of order $na^{2}$) to the value
$a\cos \phi/2$ just before the collision with scatterer $2$. We
denote the direction of the velocity of the moving particle just after
the collision with scatterer $1$ by the angle $\theta$ with respect to
some space fixed 
axis, and compute the negative Lyapunov exponent
in the following way. Since we know that the radius of curvature just
before the collision with $2$ is $
a\cos\phi
/2
$,
the radius of
curvature just after the collision with $1$ must be $
a\cos\phi
/2 +
v t $, where $t$ is the time interval between the collision with
$1$ and $2$. From this it follows that
\begin{equation}
v\int_0^{t}d\tau [\rho(\tau)]^{-1} =
\ln\left[\frac{2vt}{a\cos\phi}\right],
\label{66}
\end{equation}
where $\rho(\tau)$ is the radius of curvature of the contracting
bundle at 
times $0\leq \tau \leq t$ before the collision with $2$,
and 
$v t\gg a$.
The time average of this expression corresponds to the
result obtained by combining Eqs. (\ref{14}) and (\ref{18}),
in agreement with the conjugate pairing rule for symplectic 
systems. In the case of
a thermostatted system however, Eqs. (\ref{18}) and 
(\ref{66}) are replaced by expressions that depend on the
velocity angle $\theta$ in different ways. As a consequence
the positive and negative Lyapunov exponent are no longer each others
opposite. 

The three dimensional case proceeds in exactly the same way. We follow
an infinitesimal contracting trajectory bundle from a collision with
scatterer $1$ to a collision with scatterer $2$, such that before the
collision with scatterer $2$, the radius of curvature matrix is given
by $[(2 \cos\phi {\bf P})/a]^{-1}$ where ${\bf P}$ is defined in Eq.
 (\ref{defp}). 
We then find easily that if $t$ is the
time interval between the collisions of the moving particle with
scatterers $1$ and $2$, then the radius of
curvature matrix at some time $\tau$
between zero and $t$ 
after the
collision with scatterer $1$ is
\begin{equation}
\mbox{\boldmath$\rho$}(\tau) = v(t-\tau){\bf 1} +
\frac{a}{2\cos \phi}{\bf P}^{-1}.
\label{68}
\end{equation}
For calculating the sum of the positive Lyapunov exponents
one needs the average again of ${\rm Tr}
[\mbox{\boldmath$\rho$}(\tau)]^{-1}$, but now under the
{\it Stosszahlansatz} for the postcollisional (under the time
reversed dynamics, so in reality the precollisional) 
coordinates. However, due to the time reversal symmetry of
the dynamics the distribution of the postcollisional 
coordinates ${\phi}'$ and ${\alpha}'$ is the same as that of
$\phi$ and $\alpha$, and of course the distribution of 
intercollisional times also is the same for forward and 
backward motion. Therefore the averaging procedure yields
the same results as for Eq.\ (\ref{35}) and the sum of the negative Lyapunov exponents is given by the opposite of
Eq.\ (\ref{23}).

The individual Lyapunov exponents can be obtained using results from
the theory for eigenvalues of products of random matrices, as
described in Appendix B. The results are, as expected, that each
negative exponent is paired with a positive one such that their sum is
zero. Now we turn to the method of distribution functions for
calculating the negative Lyapunov exponents.

\subsection{The Extended Anti-Lorentz-Boltzmann Equation}

In order to use distribution functions to compute the negative
Lyapunov exponents, or their sum, we need to construct a
Boltzmann-like equation for the time reversed motion. If one reviews
the derivation of the Boltzmann equation, one sees that the colliding
particles are taken to be uncorrelated before the direct or the
restituting collisions. If we were to look at the time reversed
motion, the terms ``before" and ``after" are interchanged, and for the
time reversed motion the colliding particles are uncorrelated after
the collisions rather than before. That is, referring to Fig. 1 again,
the time reversed motion has a collision of the moving particle with
scatterer $2$, followed by a collision of the particle with scatterer
$1$. Before the collision with $2$, the moving particle is correlated
with $2$, but not so after the collision. After the collision, the
radius of curvature or the radius of curvature matrix is given by the
values 
$a\cos\phi/2$ in two dimensions, or $[
2\cos\phi{\bf
P}/a]^{-1}$ in three dimensions. The radius of curvature (2d) or the
diagonal elements of the matrix (3d) grow over
the time interval $t$ until the collision with $1$. Since the moving
particle is correlated with the scatterers before the collisions and
not after them, when deriving a Boltzmann-like equation for the
distribution function, we must apply the {\it Stosszahlansatz} to the
exiting collision cylinders after the collisions and not to the
entering cylinders before collision, as is the usual case. 

In order to clarify this procedure we first consider a derivation of
the anti-Lorentz-Boltzmann equation (ALBE) for the distribution
function $f_{-}(\vec{r},\vec{v}, t)$ for the position and velocity of
the moving particle at time $t$. We will then generalize this
derivation by adding the radius of curvature variables. Since there is
no particular problem with the time reversal of the free streaming of
the particle, we can write the ALBE in the form
\begin{equation}
\frac{\partial f_{-}(\vec{r},\vec{v},t)}{\partial t} + \vec{v}\cdot
\frac{\partial f_{-}}{\partial \vec{r}} = \Gamma_{-}^{+}-\Gamma_{-}^{-}
\label{71}
\end{equation}
where $\Gamma_{-}^{+}$ represents the rate at which particles are produced at
$\vec{r}$ with velocity $\vec{v}$ and $\Gamma_{-}^{-}$ represents the
rate at which such particles are lost. Noting the fact that we should
apply the {\it Stosszahlansatz} to the exiting collision cylinders, we
see that the rate at which particles with velocity $\vec{v}$ are
produced is
\begin{equation}
\Gamma_{-}^{+}\delta \vec{r}\delta\vec{v}\delta t = na^{d-1}\int
d\hat{n} (\hat{n}\cdot \vec{v})f_{-}(\vec{r},\vec{v},t)\delta \vec{r}\delta\vec{v}\delta t
\label{72}
\end{equation}
where $\hat{n}\cdot\vec{v}\geq 0$. That is, particles with velocity
$\vec{v}'=\vec{v}-2(\hat{n}\cdot\vec{v})\hat{n}$ collide with
scatterers and produce particles with velocity $\vec{v}$, but the
distribution function we need to calculate the rate of these
collisions
for 
is not $f_{-}(\vec{r},\vec{v}',t)$, but
$f_{-}(\vec{r},\vec{v},t)$. Similarly
\begin{equation}
\Gamma_{-}^{-} = na^{d-1}\int
d\hat{n}(\hat{n}\cdot\vec{v}')f_{-}(\vec{r},\vec{v}',t)
\label{73}
\end{equation}
where $(\hat{n}\cdot\vec{v}')\geq 0$, since particles with velocity $\vec{v}'$ are produced when a particle
of velocity $\vec{v}$ collides with a scatterer with collision vector
$\hat{n}$. Putting these terms together, we obtain
\begin{equation}
\frac{\partial f_{-}(\vec{r},\vec{v},t)}{\partial t} + \vec{v}\cdot
\frac{\partial f_{-}}{\partial \vec{r}}=na^{d-1}\int d
\hat{n}|\vec{v}\cdot\hat{n}|\left[f_{-}(\vec{r},\vec{v},t)-f_{-}(\vec{r},\vec{v}',t)\right]
\label{74}
\end{equation}
where we have used the fact that
$|\vec{v}\cdot\hat{n}|=|\vec{v}'\cdot\hat{n}|$, and the integration in
Eq. (\ref{74}) is over a semicircle (for $d=2$) or a hemisphere (for
$d=3$). This equation looks exactly like the Lorentz-Boltzmann
equation, Eq. (\ref{36}), except for the fact that the collision integral
has the opposite signs in Eqs. (\ref{74}) and (\ref{36}) \cite{Ber-Co}. 
The ALBE,
Eq. (\ref{74}) has the usual equilibrium distribution function as a
stationary solution, although it is a highly unstable solution, any
deviation will tend to grow exponentially in time. 
In fact, for this reason it is an ill-posed equation, as
arbitrarily small initial deviations can grow at arbitrarily
large rates and therefore its solution is not well-defined.
This ambiguity can be removed by requiring that the
actual solution describing a physical system after a long
enough time that all these rapidly decaying solutions
(observed in the forward time direction) have died out,
is the time reverse of the solution of the ordinary LBE (in
a closed isolated system or system coupled to a single heat 
bath this will become the equilibrium solution for long 
times). So there is little use in employing the ALBE as such,
since to obtain its physically relevant solution one has to 
solve the ordinary LBE anyway. The extension of the ALBE
discussed in the next paragraphs on the other hand does 
provide a useful tool for calculating negative Lyapunov 
exponents.

To 
this end
we need to include 
again
the radius
of curvature
matrix as variables in the distribution function,
$F_{-}(\vec{r},\vec{v},\mbox{\boldmath$\rho$},t)$. The equation for
$F_{-}$ is constructed 
as before, i. e., we write
\begin{equation}
\frac{\partial F_{-}}{\partial t}+\vec{v}\cdot\frac{\partial
F_{-}}{\partial \vec{r}}+v\sum_{i=1}^{d-1}\frac{\partial
F_{-}}{\partial \rho_{ii}} =
\Gamma_{-}^{+}(\mbox{\boldmath$\rho$})-\Gamma_{-}^{-}(\mbox{\boldmath$\rho$}).
\label{75}
\end{equation}
To compute $\Gamma_{-}^{+}(\mbox{\boldmath$\rho$})$ we multiply the
rate at which particles with velocity $\vec{v}$ are produced no matter
what their radius of curvature matrix might be,
with the fraction of
particles with velocity $\vec{v}'$ which produce particles with radius
of curvature $\mbox{\boldmath$\rho$}$ after collision. That is
\begin{equation}
\begin{array}{l}
\Gamma_{-}^{+}(\mbox{\boldmath$\rho$}) = na^{d-1}\int
d\hat{n}|\vec{v}\cdot\hat{n}|\int
d\mbox{\boldmath$\rho$}'F_{-}(\vec{r},\vec{v},\mbox{\boldmath$\rho$}',t)
\times\\
\left[\int
d\mbox{\boldmath$\rho$}''F_{-}(\vec{r},\vec{v'},\mbox{\boldmath$\rho$}'',t)\delta
(\mbox{\boldmath$\rho$}-\mbox{\boldmath$\rho$}(\mbox{\boldmath$\rho$}'')) \right]\left[\int
d\mbox{\boldmath$\rho$}''F_{-}(\vec{r},\vec{v'},\mbox{\boldmath$\rho$}'',t)\right]^{-1}.
\label{76}
\end{array}
\end{equation}
Similarly, to compute $\Gamma_{-}^{-}(\mbox{\boldmath$\rho$})$ we
multiply the rate at which particles with velocity $\vec{v}'$ are
produced due to collisions of particles of velocity $\vec{v}$ with 
scatterers by the fraction of particles of velocity $\vec{v}$ that
have the radius of curvature matrix $\mbox{\boldmath$\rho$}$. That is
\begin{equation}
\begin{array}{lcl}
\Gamma_{-}^{-}(\mbox{\boldmath$\rho$})&=&na^{d-1}\int d\hat{n}|\hat{n}\cdot
\vec{v}|\int
d\mbox{\boldmath$\rho$}'F_{-}(\vec{r},\vec{v}',\mbox{\boldmath$\rho$}',t)
F(\vec{r},\vec{v},\mbox{\boldmath$\rho$},t) \times\\
& &\left[\int
d\mbox{\boldmath$\rho$}'F_{-}(\vec{r},\vec{v},\mbox{\boldmath$\rho$}',t)\right]^{-1}.
\label{77}
\end{array}
\end{equation}
We can now assemble all of these results into an extended
anti-Lorentz-Boltzmann equation (EALBE) which reads
\begin{eqnarray}
&&\frac{\partial F_{-}}{\partial t}+\vec{v}\cdot\frac{\partial
F_{-}}{\partial \vec{r}}+v\sum_{i=1}^{d-1}\frac{\partial
F_{-}}{\partial \rho_{ii}} =  \nonumber \\
&& na^{d-1}\int
d\hat{n}|\vec{v}\cdot\hat{n}|\int
d\mbox{\boldmath$\rho$}'F_{-}(\vec{r},\vec{v},\mbox{\boldmath$\rho$}',t)
\times \nonumber \\
&&\left[\int
d\mbox{\boldmath$\rho$}''F_{-}(\vec{r},\vec{v'},\mbox{\boldmath$\rho$}'',t)\delta
(\mbox{\boldmath$\rho$}-\mbox{\boldmath$\rho$}(\mbox{\boldmath$\rho$}'')) \right]\left[\int
d\mbox{\boldmath$\rho$}''F_{-}(\vec{r},\vec{v'},\mbox{\boldmath$\rho$}'',t)\right]^{-1}-
\nonumber \\
 &&na^{d-1}\int d\hat{n}|\hat{n}\cdot\vec{v}|\int
d\mbox{\boldmath$\rho$}'F_{-}(\vec{r},\vec{v}',\mbox{\boldmath$\rho$}',t)
F(\vec{r},\vec{v},\mbox{\boldmath$\rho$},t) \times \nonumber \\
 &&\left[\int
d\mbox{\boldmath$\rho$}'F_{-}(\vec{r},\vec{v},\mbox{\boldmath$\rho$}',t)\right]^{-1}.
\label{78}
\end{eqnarray}
Here, too, $\hat{n}$ is integrated over a semicircle or a hemisphere. This 
complicated and non-linear 
looking equation (in fact one should first solve the ALBE, substitute its
solution for $\int
d\mbox{\boldmath$\rho$}'F_{-}(\vec{r},\vec{v}',\mbox{\boldmath$\rho$}',t)$
in Eq.\ (\ref{78}) and then solve the EALBE; both equations to be solved then are linear)
will form the
basis of the calculation of negative Lyapunov exponents and their sums
in the more complicated cases to be studied in subsequent
papers. 
We should note here that without further conditions this equation, like the ALBE discussed above, constitutes an ill-posed problem. The additional condition that regularizes its solution is that the integral over $\mbox{\boldmath$\rho$}$ of the distribution function yields the time reverse of the solution of the ordinary LBE. 
In the equilibrium case considered here
the equation
simplifies enormously. In fact if we use the condition that the system
is in a spatially homogeneous equilibrium state, that the integral of
the distribution function $F_{-}$ over all elements of the radius of
curvature matrix is the equilibrium distribution function $\varphi$,
which is independent of the velocity direction, we immediately obtain
Eq. (\ref{57}) for the spatially homogeneous case. Therefore for this
equilibrium situation, $F_{-}$ produces the same Lyapunov exponents as
in the forward motion, with the exception of the appropriate change of
sign due to the time reversed nature of the motion considered here.

\section{Comparison with Simulations and Discussion}

We have found that the quantities calculated in this paper, the
Lyapunov exponents and the KS entropies,
expressed as functions of the dimensionless density $\tilde{n}$, 
have
the general form
\begin{eqnarray}
\lambda_{i} = A \nu[-\ln \tilde{n} + B +
o(1)], \\
h_{KS} =  A \nu[-\ln \tilde{n} + B +
o(1)],
\end{eqnarray}
where $\nu$ and $\tilde{n}$ are the collision frequency and reduced
density,
given by $\nu = 2nav, \tilde{n}=na^{2}$ in two dimensions and
$\nu =\pi na^{2}v, \tilde{n}=\pi na^{3}$ in three dimensions, and
$A,B$ are constants which we have determined. In Table I we compare
the
theoretical results for $A
$
and $B$ with values for the same
coefficients,
as obtained by Dellago and Posch\cite{Pos-Del} from computer
simulations of two and three dimensional hard sphere Lorentz
gases. The results are in excellent agreement for the coefficient $A$
and there are minor discrepancies in the $B$ values, probably due to
the fact that the simulation analysis is difficult at the low
densities where the theoretical analysis given here applies.

It is remarkable that the Lyapunov exponents for the 3 dimensional completely 
isotropic random Lorentz gas are different at all. In fact,
on the basis of the
results
obtained in leading order in the density,
it 
has been conjectured 
\cite{Chernov} 
that 
all
positive 
as well as all
negative Lyapunov exponents are 
equal.
The methods used
in our approach allow  a very transparent explanation for the 
differences, 
now also 
confirmed
numerically, 
between
the Lyapunov exponents. Especially it 
becomes
clear why all Lyapunov exponents coincide in leading order but differ in next
to leading order.  The reason
for this
is
related to the different nature of the terms
contributing to the different orders \cite{latz}. 
The terms proportional to $n \log( n/2) + {\cal C}$ in Eqs. (\ref{lammax}, 
\ref{24}) 
result from
averaging over functions, which only depend on the time 
of free flight. Due to the isotropy 
of the free flight this has to lead to 
equal
Lyapunov exponents. The 
{\it differences} arise
on averaging over functions, which depend on the 
collision parameters $\phi$ and $ \alpha$. To understand how the  
scattering process, which is 
isotropic 
for a single
trajectory, can cause the Lyapunov exponents to be different, it is worthwhile 
considering 
scattering events in more detail. 

For calculating the Lyapunov exponents we have to analyze the scattering of 
two close by trajectories. Therefore it becomes possible for a given 
scattering angle $\phi$ to distinguish whether $\delta \vec{r}_\perp$ is in 
the plane $\cal P$ 
spanned by the normal $\hat{n}$ on the sphere and the 
impact velocity $\vec{v}$ or perpendicular to $\cal P$. In this way the isotropy of 
the scattering process becomes effectively broken. This is reflected in the 
eigenvalue structure of the radius of curvature matrix 
$\mbox{\boldmath$\rho^+$} = (2 \cos \phi 
{\bf P}/a)^{-1}$
(see Eq. (\ref{defp})). There are two eigenvalues $\rho_1 = 
a \cos \phi /2$ 
and $\rho_2 = 
a/(2 \cos \phi)$.  Here the eigenvalue $\rho_1$ 
corresponds
to the
 eigendirection $\vec{e}_1$ which 
 is in the plane $\cal P$, and the eigenvalue
$\rho_{2}$  corresponds to the eigendirection which is perpendicular to
 the plane $\cal P$, with both eigendirections perpendicular to $\vec{v}$. 
Note that for $\phi =0$ the eigenvalues are the same. This can be
understood by realizing that for $\phi =0$, the particle hits the sphere head on , i.e 
$\vec{v}$ is parallel to $\hat n$. For this special case the two
eigendirections are clearly equivalent, but for other values of
$\phi$, this symmetry is lost, and the eigenvalues differ. Therefore
we may conclude that the lack of degeneracy of the positive Lyapunov
exponents is due to the lack of rotational symmetry when the nearby
trajectories hit the sphere.  

We conclude with a number of remarks:

1) The results given here can be extended to higher densities in a
number of ways. In particular, BBGKY hierarchy methods are being
developed to provide a systematic density expansion of the Lyapunov
exponents and the KS entropies beyond the low density results obtained
here. This will be especially important when non equilibrium situations
are considered, since there one 
may 
see the effects of long time tail
phenomena on the chaotic properties of the system. 

2) As remarked earlier, we relied upon the low dimensionality of our
systems to obtain all of the relevant Lyapunov exponents. For a four
dimensional Lorentz gas we would need to use more sophisticated
techniques to obtain all of the Lyapunov exponents, since the methods
given here could only provide values for the largest exponent and the
sum of all of three of the positive exponents. We could not then resolve
the two smaller positive exponents, but only could get their sum.

3) We have not analyzed here a particularly interesting quantity that
gives a more general characterization of the chaotic properties of the
system, namely the Ruelle, or topological, pressure\cite{Ru,Be-Sch}. This quantity has
the formal structure of an equilibrium free energy and it depends upon
a temperature-like parameter, $\beta$. The results obtained here
characterize the chaotic properties in the neighborhood of $\beta=1$. 
Elsewhere it has been shown that for the
Lorentz gas, where the disorder is static,
in the thermodynamic limit the topological pressure
exhibits a localization transition as a function of $\beta$,
it 
is dominated by
contributions from particles which are localized within the 
largest dense cluster for
$\beta <1$, and in the largest 
region without scatterers for $\beta >1$\cite{AEVBD}. 
It would
be very valuable to obtain these results using kinetic theory methods
in addition to the more rigorous analysis given in Ref.\ 
\cite{AEVBD}.
 
4) It should be straightforward to extend the results obtained here to
other potentials of interaction between the moving particle and the
fixed scatterers, at least for dilute systems. It is well known that
the Boltzmann transport equation can be applied to gases that interact
with other than hard core potentials \cite{Chap-Cow}. In fact a wide
variety of interaction potentials may be used in the Boltzmann
equation to determine the transport properties of the corresponding
gases. Certainly Lyapunov exponents can be calculated for Lorentz gases where the moving particle has other than
hard core interactions with the scatterers, as well.

5) Finally we mention that the method given here can be adapted to 
cases
where all the particles in the system are moving, namely, a
dilute gas. Expressions for the KS entropy\cite{VBDDP} 
and the largest Lyapunov
exponent of two and three dimensional gases with short range forces\cite{rh}
have already been obtained this way.

In the next paper we will extend the results here to open systems with
escape and compute  the escape rates, Lyapunov exponents, and KS
entropies characterizing the fractal repeller that underlies diffusion
in open systems with absorbing boundaries. This work will make heavy
use of the extended LB equation, and we regard the
present paper as an introduction to the next one.
\newpage
\begin{table}
\begin{tabular}{p{3cm}|p{3cm}|p{3cm}}
QUANTITY & THEORY & SIMUL. \\ \hline
2d: $\lambda^{+}$ & \ & \\ \hline
A & 1 & 0.995 $\pm$ 0.009 \\
B & 0.423 & 0.463 $\pm$ 0.083 \\ \hline
3d: $\lambda^{+}_{max}$ & \ & \\ \hline
A & 1   & 0.990 $\pm$ 0.089\\
B & 0.309 & 0.387 $\pm$ 0.746 \\ \hline
3d: $\lambda^{+}_{min}$ & \ & \\ \hline
A  & 1 &  0.992 $\pm$ 0.084 \\
B & -0.077 & -0.015 $\pm$ 0.715 \\ \hline
3d: $h_{KS}$ & \ &         \\ \hline
A & 2 & 1.982 $\pm$ 0.173 \\
B & 0.166 & 0.372 $\pm$ 1.461  \\ \hline
\end{tabular} 
\caption{ Comparison of Theoretical and Simulation Results}
\end{table}

ACKNOWLEDGEMENTS
The authors would like to thank  E. G. D. Cohen, Ch. Dellago,
M. H. Ernst, H. A. Posch, C. Appert, and R. van Zon for many helpful
conversations as this work progressed. Ch. Dellago and H. A. Posch
kindly supplied the results of their simulations which were used in
Table I. We thank them as well as C. Ferguson who carried out the
analysis of their data. J.R.D wishes
to thank D. Panja for useful
conversations as well as the National Science Foundation for support
under grants PHY-93-21312, and PHY-96-00428. A.L. thanks the DFG, 
through SFB 262, for financial support during the time this paper was 
written up.  H.v.B
was supported by FOM, SMC and by the NWO Priority Program Non-Linear
Systems, which are financially supported by the "Nederlandse Organisatie voor Wetenschappelijk Onderzoek (NWO)".  He thanks the ENS-Lyon for its hospitality
during part of this project.

\appendix

\section{Binary Collision Dynamics and the radius of Curvature
Matrix}

Here we give a brief review of the derivation of the formulae for the
change in the spatial and velocity deviations at a binary collision used in 
Eqs. (\ref{4a}, \ref{4}) based on the
method of Dellago, Posch and Hoover\cite{DePoHo}. We relate these formulas
to
the expression used for the change in the radius of curvature matrix,
given by Eq. (\ref{30}), which, in turn, was discussed by Gaspard and
Dorfman\cite{Gas-Do}. Consider a trajectory of the particle moving among the
scatterers. We denote the initial position and velocity of this
trajectory by $x(0)$, and the position and velocity at time $t$ later
by $x(t)$. Consider also a trajectory that is obtained by an
infinitesimal displacement of the initial position and velocity to
$x(0)+\delta x(0)$, and denote the position and velocity of the second
trajectory at time $t$ by $x(t)+\delta x(t)$. We 
require the two
trajectories 
to be 
infinitesimally close. 
Our goal is 
deriving equations for $\delta x(t) = (\delta
\vec{r}(t),\delta \vec{v}(t))$.  
We take all trajectories 
with the same energy,
which leads to the
condition that $\delta \vec{v}(t)\cdot \vec{v}(t)=0$, so the velocity
deviation is always perpendicular to the velocity $\vec{
v}(t)$. We can
also set the position deviation $\delta \vec{r}(t)\cdot \vec{v}(t)=0$,
since this simply requires that the position deviation is
perpendicular to the velocity at the initial time. 

Now in between
collisions $\delta \vec{r}(t), \delta \vec{v}(t)$ satisfy
Eqs. (\ref{3a}, \ref{3}). However the change in these quantities at collision is
more complicated. To analyze this change we suppose that the
trajectory with $x(t)$ has a collision with some scatterer at time
$\tau$. Then immediately after the collision the velocity has changed
to $\vec{v}^+ =\vec{v}-2(\vec{v}\cdot \hat{n})\hat{n}$ where $\hat{n}$
is a unit vector in the direction from the center of the scatterer to
the point of contact, and $\vec{v}$ is the velocity immediately before
collision. The displaced trajectory will have a collision at a
slightly displaced time $\tau + \delta \tau$, and at a slightly
different point on the same scatterer located by unit vector $\hat{n}+\delta
\hat{n}$, with $\hat{n}\cdot\delta\hat{n}=0$. By examining the scattering equations for the displaced
trajectory one easily finds
\begin{eqnarray}
\delta\vec{v}^+ = ({\bf{1}}-2\hat{n}\hat{n})\cdot\delta\vec{v}
-2[(\vec{v}\cdot \hat{n})\delta\hat{n}+(\vec{v}\cdot\delta\hat{n})
\hat{n}] \\
\delta\vec{r}(\tau)+\delta\tau \vec{v} = a\delta\hat{n} \\
\delta \vec{r}^+ = 
-
\delta\tau \vec{v}^+ 
+
 a\delta\hat{n}.
\end{eqnarray}
We use the condition $\hat{n}\cdot\delta\hat{n}=0$ to obtain $\delta
\tau = -(\hat{n}\cdot\delta\vec{r}(\tau))/(\vec{v}\cdot\hat{n})$. 
Simple
algebra leads to Eqs. (\ref{4a}, \ref{4}) in the text. It is important to
note that $\delta\vec{r}(\tau)$ is the spatial deviation when the
"main" trajectory has a collision, while $\delta\vec{r}^+$ is the
spatial separation at the instant that the displaced trajectory has a
collision. (It is easy to visualize this in the case that $\delta \tau
>0$, as illustrated in Fig. 2.)

Now we introduce the radius of curvature matrix as given by
Eq. (\ref{28}). We note that 
$\mbox{\boldmath$\rho$}$ is a matrix of rank $d-1$.
If we
substitute the definition of this matrix, Eq. (\ref{28}) into Eq. (\ref{4a}),
 we obtain
\begin{equation}
\mbox{\boldmath$\rho$}^{+}\cdot\delta\vec{v}^{+}=
{\bf {U}}\cdot\mbox{\boldmath$\rho$}^{-}\cdot \delta\vec{v}^{-},
\label{a1}
\end{equation}
where ${\bf {U}} ={\bf {U}}^{-1}={\bf{1}}-2\hat{n}\hat{n}$ is 
a reflection matrix and is discussed in
Ref. \cite{Gas-Do}. Here the superscripts $+,-$ denote after 
and before
collision respectively. The operators ${\bf U}$ have 
determinant
-1
and they ensure the proper
orientations, at collision, of the planes in which the radius of curvature
matrices are defined. We now substitute Eq. (\ref{4}) for
$\delta\vec{v}^{+}$ into Eq. (\ref{a1}), use the relation Eq. (\ref{28}), 
and
some elementary matrix manipulations to obtain Eq. (\ref{30}).
As mentioned in the main text the inverse matrices 
$[\mbox{\boldmath $\rho$}_{u}^{-1(+)}]$ and $[\mbox{\boldmath $\rho$}_{u}^{-1(-)}]$ are defined in the subspaces orthogonal
to $\vec{v}$ and ${\vec{v}}'$ respectively.

We can obtain a considerable simplification of the analysis of the spatial deviation
vector, $\delta \vec{r}_{\perp}^{(-)}$ in Eq. (\ref{32}) by using the
properties of the  matrices
${\bf U}$. That is, we can easily show that it is possible to
express the spatial deviation vector in terms of radius of curvature
matrices $\bar{\mbox{\boldmath $\rho$}}$ all of which are defined in a
plane perpendicular to the initial velocity $\vec{v}(0)$. To see this,
consider the right hand side of Eq. (\ref{32}) for the case that $n=2$. We
have
\begin{equation}
\begin{array}{lcl}
\delta \vec{r}_{\perp}^{(-)}(t_2) & = & [{\bf {1}}_{\perp}(1) +
v\tau_{2,1}{\mbox{\boldmath $\rho$}_{u}^{-1}}^{(+)}(t_1)]\cdot {\bf
U}(1)\cdot \\
& &[{\bf 1}_{\perp}(0) + v\tau_{1,0}\mbox{\boldmath
$\rho$}_{u}^{-1}(0)]\cdot \delta \vec{r}_{\perp}(0).
\end{array}
\label{a2}
\end{equation}
Here ${\bf 1}_{\perp}(i)$ is a unit operator in the plane
perpendicular to the velocity vector after the $i-th$ collision. Note that 
Eq. (\ref{30}) allows us to write 
\begin{equation}
{{\mbox{\boldmath $\rho$}_{u}^{-1(+)}}}(t_1) ={\bf
U}(1)\cdot{{\bar{\mbox{\boldmath
$\rho$}}_{u}^{-1(+)}}}(t_1)\cdot{\bf U}(1),
\label{a3}
\end{equation}
where
\begin{equation}
{{\bar{\mbox{\boldmath $\rho$}}_{u}^{-1(+)}}}(t_1) = 
\mbox{\boldmath
$\rho$}_{u}^{-1}(0) +\frac{2}{a}\left[ \hat{v}(0)\hat{n}_1 + \hat{n}_1
\hat{v}(0) - \frac{1}{(\hat{v}(0)\cdot\hat{n}_1)}\hat{n}_1\hat{n}_1
-(\hat{v}(0)\cdot\hat{n}_1){\bf 1}\right].
\label{a4}
\end{equation}
It is important to note that the operator defined by the square
bracket appearing on the
right hand side of Eq. (\ref{a4}) is orthogonal from the right and
left to $\vec{v}(0)$, as is $\mbox{\boldmath $\rho$}_{u}$. Therefore
the operator $\bar{\mbox{\boldmath $\rho$}}_{u}^{(+)}$ is to be
evaluated in the plane perpendicular to $\vec{v}(0)$. Furthermore the unit 
operator
${\bf{1}}_{\perp}(1)$ appearing in the same bracket 
with ${\mbox{\boldmath $\rho$}_{u}^{-1}}^{(+)}(t_1)$ 
can be written as ${\bf
U}(1)\cdot{\bf 1}_{\perp}(0)\cdot {\bf U}(1)$. From this it
follows that 
\begin{equation}
\delta \vec{r}_{\perp}^{(-)}(t_2) = {\bf U}(1)\cdot[{\bf
1}_{\perp}(0)+v\tau_{2,1}{{\bar{\mbox{\boldmath
$\rho$}}_{u}^{-1(+)}}}(t_1)]\cdot [{\bf 1}_{\perp}(0)+\mbox{\boldmath $\rho$}_{u}^{-1}(0)]\cdot
\delta \vec{r}_{\perp}(0).
\end{equation}
If this procedure is followed through each successive collision,
Eq. (\ref{32}) can easily be written as 
\begin{equation}
\delta \vec{r}_{\perp}^{(-)}(t_n)={\bf U}(n-1)\cdot{\bf
U}(n-2)\cdots {\bf U}(1)\cdot
\bar{\delta\vec{r}}_{\perp}^{(-)}(t_n),
\end{equation}
where 
\begin{equation}
\bar{\delta\vec{r}}_{\perp}^{(-)}(t_n) = [{\bf
1}_{\perp}(0)+{{\bar{\mbox{\boldmath
$\rho$}}_{u}^{-1(+)}}}(t_{n-1})]\cdots [{\bf 1}_{\perp}(0) + \mbox{\boldmath $\rho$}_{u}^{-1}(0)]\cdot
\delta \vec{r}_{\perp}(0).
\end{equation}
The product of the ${\bf U}$ matrices appearing on the right
hand side of Eq. (A9) have determinant 
$\pm 1$, of course, and have
no bearing on the exponential growth of the spatial deviation
vector. As a result one can carry out all calculations of the Lyapunov
exponents and the KS entropy in a coordinate system defined 
in the plane perpendicular to
the
initial velocity $\vec{v}(0)$.
As a result, all of the ${\bf U}$ operators can be dispensed
with in the calculation of the Lyapunov exponents provided one uses
$\bar{\mbox{\boldmath $\rho$}}_{u}$ operators as well as unit
operators ${\bf 1}_{\perp}(0)$. This procedure ``unwinds" the
trajectory of the moving particle and allows all collisions with the
scatterers to be treated in one coordinate system. This is useful when
one wants to avoid neglecting the left hand side of Eq. (\ref{44}), to treat
systems at higher densities, and/or to use the methods of random
matrix theory in a convenient way, as we do in the next appendix. 

\section{The Solution of the Extended LB Equation Using the Method of
Characteristics, and Random Matrix Methods}

Here we indicate how Eq. (\ref{57}) can be solved as a differential equation
in three variables using the method of characteristics\cite{Cou-Hil}. We begin with
Eqs. (\ref{60}), and change variables to $\sigma_1=2\rho_1/a,
\sigma_2=2\rho_2/a, \sigma_{12}=2\rho_{12}/a$. We also set
$g(\rho_1,\rho_2,\rho_{12}) = 8/a^3 G(\sigma_1,\sigma_2,\sigma_{12})$ and 
introduce the scaled density $\tilde{n} = n \pi a^3$. Then
the equation for $G$ is
\begin{equation} \label{c1}
\begin{array}{l}
\left(\frac{\partial}{\partial \sigma_1}+\frac{\partial}{\partial
\sigma_2}\right)G +\frac{\tilde{n}}{2}G = \\ 
4 K \Theta(1-\sigma_1)\Theta
\left(1-\frac{2|\sigma_{12}|}{|\sigma_2
-\sigma_1|}\right)\left(\frac{\sigma_{1}^{2}}{(1-\sigma_{1}^{2})}\right)\delta(\sigma_2 -
\frac{1}{\sigma_1}).
\end{array}
\end{equation}
where $K= \frac{\tilde{n}}{2 \pi}$. To use the
method of characteristics we write 
\begin{equation}
\left(\frac{\partial}{\partial \sigma_1}+\frac{\partial}{\partial
\sigma_2}\right)G = \frac{d}{ds}G
\end{equation}
where
\begin{equation}\label{chlines}
\begin{array}{l}
\frac{d}{ds}\sigma_1 = 1; \,\, \sigma_1 = \sigma_{1}^0 +s  \\
\frac{d}{ds}\sigma_2 = 1; \,\, \sigma_2 = \sigma_{2}^0 +s
\\
\frac{d}{ds}\sigma_{12}=0; \,\, \sigma_{12}=\sigma_{12}^{0}. 
\end{array}.
\end{equation}
This substitution converts the partial differential equation into a
simple differential equation which can be solved by elementary means
as a function of $s$, once we have specified the appropriate boundary
conditions. Further, in the $\sigma_1,\sigma_2$ plane, the $s$
integration  corresponds to an integration along a set of lines given
by Eqs. (\ref{chlines}), called the characteristic lines.
The inhomogeneous term on the right hand side of
Eq. (\ref{c1}) is zero everywhere except on the line
$\sigma_2 = 1/\sigma_1;
\  \sigma_1 \leq 1$. We look for
solutions that vanish at $\sigma_1 =0$
and at $\sigma_2 =0$, and we
note that the solution $\hat{G}$ of the homogeneous equation,
expressed in terms of $s$, has the
form
\begin{equation}
\hat{G}=G_0 e^{-\frac{\tilde{n}s}{2}}.
\label{c2}
\end{equation}
All of the conditions can be satisfied if $G$ vanishes in the
$\sigma_
1
,\sigma_2$ plane, except in the region defined by the curves $\sigma_2
\geq \sigma_1$, $\sigma_1 > 1$ and $\sigma_2 \geq 1/\sigma_1, \sigma_1 
\leq 1$. See
Fig. 3. Choosing $\sigma_1^0 = z \le 1$,$ \;$  $\sigma_2^0 = 1/\sigma_1^0 $ and
$\sigma_{12} = \sigma_{12}^0: 
= \sin (2 \hat{\alpha}) \,  
|\sigma_2-\sigma_1|/2$ 
the characteristic lines are given by 
\begin{equation} \label{c3}
\begin{array}{l}
\sigma_1 = z + s ; \,\, z \leq 1 \\
\sigma_2 =\frac{1}{z} +s.\\
\sigma_{12} = \frac{1 -z^2}{2z} \sin(2 \hat{\alpha})
\end{array}
\end{equation}
We then arrive at the equation
\begin{equation}
\frac{d}{ds}G + \frac{\tilde{n}}{2} G =
K\Theta(1-z)
\Theta(2 \pi - \hat{\alpha}) \Theta(\hat{\alpha})\left( \frac{z^2}{1-z^2}\right)\left(
\frac{z^2}{1+z^2}\right) \delta(s).
\label{c4}
\end{equation}
The probability density for the random variables $s,z,\hat{\alpha}$ is given by
$\hat{f}(s,z,\hat{\alpha}) = J(s,z,\hat{\alpha}) G(s,z,\hat{\alpha})$ 
where $J$ 
is the Jacobian $J= |\partial (\sigma_1,\sigma_2,\sigma_{12})/\partial
(s,z,\hat{\alpha})| = 
(1-z^2)(1+z^2)/z^3$. $\hat{f}$ obeys the equation 
\begin{equation} \label{f1}
\frac{d}{ds} \hat{f} + \frac{\tilde{n}}{2} \hat{f} = K\Theta(1-z)
\Theta(2 \pi -\hat{\alpha}) \Theta(\hat{\alpha}) \, z \, \delta(s).
\end{equation}
This equation can now be solved along the characteristic lines in the
region indicated in Fig. 3. It has the simple solution 
\begin{equation} \label{f2}
\hat{f} = \frac{\tilde{n}}{2 \pi} \Theta(1 -z) \Theta(s) \Theta(\hat{\alpha}) 
\Theta(2 \pi - \hat{\alpha}) z \exp (- \frac{\tilde{n}}{2} s).
\end{equation}
When the solution is inserted in the
expression for $h_{KS} = v \langle Tr {\bf I}/\mbox{\boldmath$\rho$} \rangle$,
\begin{equation}
h_{KS} = v \int d \rho_1 d \rho_2 d \rho_{12} g(\rho_1, \rho_2, \rho_{12})
(\frac{1}{\rho_1} + \frac{1}{\rho_2}) = \frac{2 v}{a} \int ds dz d\hat{\alpha}
\hat{f}(s,z,\hat{\alpha}) (\frac{1}{z +s} + \frac{1}{1/z+s}),
\end{equation}
the results obtained agree with those given by Eq. (\ref{23}).

The calculation of the maximum Lyapunov exponent starts from the observation 
that the deviations $\delta \vec{r}_\perp (t)$ are given by a product of random
 matrices acting on the initial deviation $\delta \vec{r}_\perp (0)$ (see 
Eq. (\ref{35d})). The random matrix ${\bf R}(t_{j+1},t_j)$, Eq. (\ref{35e}), 
depends on the 
time of free flight between two collisions and the initial matrix $\rho^+$ 
given by $((a\/(2 \cos\phi)){\bf P}^{-1}$. ${\bf P}$ (see. Eq.(\ref{defp})),
 is  parameterized by two collision 
angles 
$\phi$, $\alpha$. Another way of parameterizing these initial matrix is by 
using its random eigenvalues $z:= \sigma_1^0  = \cos \phi$, $\sigma_2^0=1/z$ 
and the the random angle variable $\alpha$. Notice that the characteristic 
lines are identical to the free flight solutions for the eigenvalues of $(2/a)
 \mbox{\boldmath$\rho$}$ and its off diagonal element $\sigma_{12}$
(Compare eq. 
(\ref{chlines})). They are chosen such, that the initial conditions of the  
characteristic lines ($s=0$) are the values of $ \sigma_i,\sigma_{12}$ 
immediately after a scattering event with scattering angles $\phi,\alpha$,
Therefore
(\ref{f2}) can be interpreted as  
the distribution function for the random variables $z =\cos \phi$ 
(cosine of the polar scattering angle), $\alpha$ (azimuthal scattering angle) 
and the time of free flight $s$.  

The calculation of the maximum eigenvalue is now a straight forward 
application of the theory of products of random matrices \cite{Vulp}. 
It follows from 
Eq. (\ref{35d}) that the maximum Lyapunov exponent is given by 
\begin{equation} \label{c5}
\lim_{t->\infty} \frac{1}{t} \log |{\bf {\Pi}R} \delta \vec{r}(0)|/|\delta 
\vec{r}(0)|
\end{equation}
with ${\bf {\Pi}R} = \prod_{j=0}^\infty {\bf R}(j,j+1)$. Using an analogous 
decomposition of the product as in Eq. (\ref{9}) and  
the identity $t = \sum_{i=1}^N s_i$, valid right 
after 
the Nth collision,  
with $s_i$ 
the 
time of free 
flight between 
collisions
$i-1$ and $i$, (\ref{c5}) is equivalent to
 
\begin{equation} \label{c6}
\lambda_{max} = \frac{1}{\langle s \rangle } 
\overline{\langle \log |{\bf R}(z,\alpha,s)\cdot 
\vec{e}_\psi| \rangle}^{\psi}. 
\end{equation}
Here
$\vec{e}_\psi$ is a unit vector $\vec{e}_\psi = (\cos \psi, 
\sin \psi)$
and
$\overline{\langle \dots \rangle}^{\psi}$ indicates an average over the 
distribution of $(z,\alpha,s)$ plus an additional average over a stationary 
distribution $D(\psi)$ of directions  $\vec{e}_\psi$. 
This distribution is a 
solution of a 
Frobenius-Perron equation. 
\begin{equation}\label{perron}
D(\psi) = \int^{2 \pi}_0 d\psi' D(\psi') \delta(\psi' - 
\cos^{-1}(\frac{\vec{e}_x\cdot {\bf R}(z,\alpha,s)\cdot 
\vec{e}_\psi}{|\bf R(z,\alpha,s)\cdot 
\vec{e}_\psi|})
\end{equation}
As we will show shortly this additional average is not necessary in the 
equilibrium case since $\psi$ can be absorbed in a redefinition of 
the azimuthal scattering angle $\alpha$.    
The equation for $\bf R$ can be derived from the equation (\ref{35e}).
\begin{equation}\label{def_u}
\frac{d {\bf R}}{ds} = v \mbox{\boldmath$\rho$}^{-1}{\bf R}
\end{equation}
In the equilibrium this is easily solved using \mbox{\boldmath$\rho$}(s)  = 
\mbox{\boldmath$\rho^+$} + v {\bf 1} s
\begin{equation} \label{c7}
{\bf R}(z,\alpha,s) = {\bf 1} + \mbox{\boldmath$\rho^+$}^{-1} s \approx 
\frac{2 \cos \phi}{a} {\bf P} \, s.
\end{equation}
The last approximation is valid in the low density limit, 
since the most important contributions to the 
average over $s$ come from large time of free flight $s$. ${\bf P}$ is 
defined in Eq. (\ref{defp}). With this we obtain  
\begin{equation}\label{c8}
|{\bf R}(z,\alpha,s)\cdot \vec{e}_\psi| =  \sqrt{(\frac{1+z^2}{z})^2 + 
(\frac{1-z^2}{z})^2 + 2 \frac{(1+z^2)(1-z^2)}{z^2 } 
\cos ( 2( \alpha - \psi))} s
\end{equation}
From Eq. (\ref{c8}) it is obvious that $|{\bf R}(z,\alpha,s)\cdot \vec{e}_\psi|$
 is statistically independent of $\psi$ i.e. the dependence on $\psi$
 can be absorbed in a redefinition of $\alpha$.
 To obtain $\lambda_{max}$ we have to calculate the average of Eq.(\ref{c8})
using the distribution function Eq. (\ref{f2}). 
The result agrees with Eq. (\ref{lammax}).

\newpage 
\centerline{\bf FIGURE CAPTIONS}

{\it Figure 1}. A trajectory from scatterer $1$ to scatterer $2$ with bundles of
expanding and contracting trajectories indicated.

\vskip 0.5 cm

{\it Figure 2}. The arrangement of the position and velocity vectors for one
trajectory at the instant of a collision
of the moving particle with the scatterer and at the instant of the
collision of the particle with the same scatterer along an infinitesimally displaced
trajectory.
    
\vskip 0.5 cm

{\it Figure 3}. The solution regions for Eq. (\ref{c1}) using the method of
characteristics. A characteristic line is indicated parallel to the
line $\sigma_{1}=\sigma_{2}$.


\begin{references}

\bibitem{Hauge} E.H. Hauge, in {\it Transport Phenomena}, Lecture Notes
in Physics, vol. 31, edited by G. Kirczenow and J. Marrow, (Springer
Verlag, Berlin, 1974), p.337.

\bibitem{Cohen1} E. G. D. Cohen, Colloq. Int. (CNRS), {\bf 236}, 269, (1974).

\bibitem{Sinai1} Ya. G. Sinai, Russ. Math. Surv. {\bf 25}, 137, (1970);
L. A. Bunimovich and Ya. G. Sinai, Comm. Math. Phys. {\bf 78}, 479,
(1981); Ya. G. Sinai and N. I. Chernov, Russ. Math. Surv. {\bf 42}, 181,
(1987); L. A. Bunimovich in {\it Dynamical Systems II} edited by
Ya. G. Sinai, (Springer Verlag, Berlin, 1989) p. 151; P. Gaspard in
{\it Quantum Chaos}, edited by G. Casati, I. Guarneri, and
U. Smilansky (North Holland, Amsterdam, 1993), p.307. 


\bibitem{Sinai2} Ya. G. Sinai, Funct. Anal. Appl. {\bf 13}, 192, (1980).

\bibitem{Gal-Or} G. Gallavotti and D. S. Ornstein,
Comm. Math. Phys. {\bf 38}, 83, (1974).

\bibitem{Che-Has} N. I. Chernov and C. Haskell, Erg. Th. and
Dyn. Syst. {\bf 16}, 19, (1996). 

\bibitem{Krylov} N.S. Krylov, {\it Works on the Foundations of
Statistical Physics}, edited by Ya. G. Sinai, (Princeton University
Press, Princeton, 1979).

\bibitem{Chernov} N. I. Chernov, Funct. Anal. Appl. {\bf 25}, 204, (1991).

\bibitem{Ru-Eck} J.-P. Eckmann and D. Ruelle, Rev. Mod. Phys. {\bf 57},
617, (1985).

\bibitem{hvb1} H. van Beijeren and J. R. Dorfman,
Phys. Rev. Lett. {\bf 74}, 4412, (1995).

\bibitem{hvb2} H. van Beijeren, J. R. Dorfman, E. G. D. Cohen,
H. A. Posch, and Ch. Dellago, Phys. Rev. Lett. {\bf 77}, 1974, (1996). 

\bibitem{latz1}  A. Latz, H. van Beijeren, and J. R. Dorfman,
Phys. Rev. Lett. {\bf 78}, 207, (1997). see also J. R. Dorfman and
H. van Beijeren, Physica A, {\bf 240}, 12, (1997).

\bibitem{latz} A. Latz, Z. Phys. B {\bf 103}, 351 (1997).

\bibitem{Gas-Nic} P. Gaspard and G. Nicolis, Phys. Rev. Lett. {\bf 65},
1693, (1990).

\bibitem{Evans-Hoo} D. J. Evans and G. P. Morriss, {\it Statistical
Mechanics of Nonequilibrium Liquids}, (Academic Press, London, 1990);
W. G. Hoover, {\it Computational Statistical Mechanics}, (Elsevier,
Amsterdam, 1991).



\bibitem{Pos-Del} Ch. Dellago and H. A. Posch, Phys. Rev. E {\bf 52},
2401, (1995); Phys. Rev. Lett. {\bf 78}, 211, (1997).

\bibitem{Chap-Cow} S. Chapman and T. G. Cowling, {\it Mathematical
Theory of Non-Uniform Gases}, 
(Cambridge University Press,
Cambridge, 1970). 

\bibitem{Arnold} V. I. Arnold, {\it Mathematical Methods of Classical
Mechanics}, 
(Springer Verlag, Berlin, 1989). 

\bibitem{Gas-Do} P. Gaspard and J. R. Dorfman, Phys. Rev. E {\bf 52},
3525, (1995).

\bibitem{DePoHo} Ch. Dellago, H. A. Posch, and W. G. Hoover,
Phys. Rev. E {\bf 53}, 1485, (1996).

\bibitem{Flanders} H. Flanders, {\it Differential Forms with
Applications to the Physical Sciences}, (Dover Publ. Co, New York, 1989). 

\bibitem{Vattay} G. Vattay, Prog. Theor. Phys. Suppl. {\bf 116}, 251, (1994).

\bibitem{Vulp}D. Mannion, Ann. Appl. Prob. {\bf 3}, 1189 (1993); 
A. Crisanti, G. Paladin, and A. Vulpiani, {\it Products
of Random Matrices in Statistical Physics}, (Springer Verlag, Berlin, 1993).

\bibitem{Cohen2} J. R. Dorfman and E. 
G. D. Cohen, J. Math. Phys. {\bf
8}, 282, (1967).

\bibitem{Do-HVB} J. R. Dorfman and H. van Beijeren, in {\it Statistical
Mechanics, Part B}, edited by B. J. Berne, (Plenum Press, New York, 1977).

\bibitem{hvl-weij} A. Weijland and J. M. J. van Leeuwen, Physica {\bf
36}, 457, (1967); {\bf 38}, 3, (1968); see also C. Bruin, Physica {\bf
72}, 261, (1974).

\bibitem{ern-weij} M. H. Ernst and A. Weijland, Physics Lett. A {\bf
34}, 39, (1971).

\bibitem{EvMoCo} D. J. Evans, E. G. D. Cohen, and G. P. Morriss,
Phys. Rev. A {\bf 42}, 5990, (1990). 

\bibitem{DettMo} C. P. Dettmann and G. P. Morriss, Phys. Rev. E {\bf
 53}, R5545, (1996).

\bibitem{Woj-Liv} M. P. Wojtkowski and C. Liverani (to be published).

\bibitem{Ber-Co} E. G. D. Cohen and T. Berlin, Physica, {\bf 26}, 717, (1960).



\bibitem{Ru} D. Ruelle, {\it Thermodynamic Formalism} (Addison Wesley,
Reading, MA, 1978).

\bibitem{Be-Sch} C. Beck and F. Schl\"ogl, {\it Thermodynamics of
Chaotic Systems}, (Cambridge University Press, Cambridge, 1993).

\bibitem{AEVBD} C. Appert, H. van Beijeren, M. H. Ernst, and
J. R. Dorfman, Phys. Rev. E {\bf 54}, R1013, (1996);
J. Stat. Phys. {\bf 87}, 1253, (1997).

\bibitem{VBDDP} H. van Beijeren, J. R. Dorfman, Ch. Dellago, and
H. A. Posch, Phys. Rev. E (to appear)
[chao-dyn 9706019].

\bibitem{rh} R.\ van Zon and H.\ van Beijeren, submitted to Phys.\
Rev.\ Lett.\ [chao-dyn ???]
\bibitem{Cou-Hil} E. Courant and D. Hilbert, {\it Methods of
Mathematical Physics}, Vol II, (John Wiley-Interscience Publ. Co., New York, 1962).

\end{references}
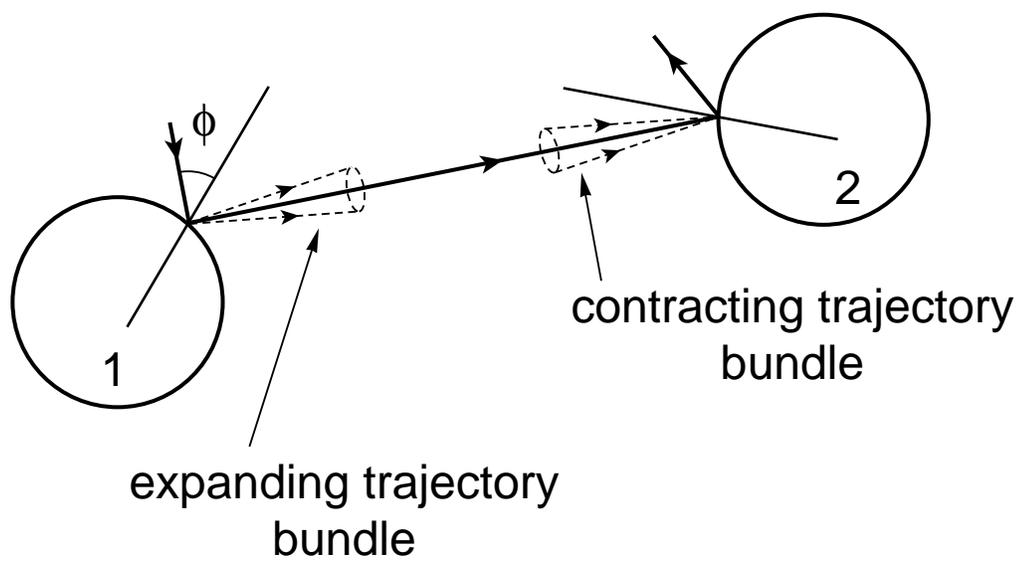
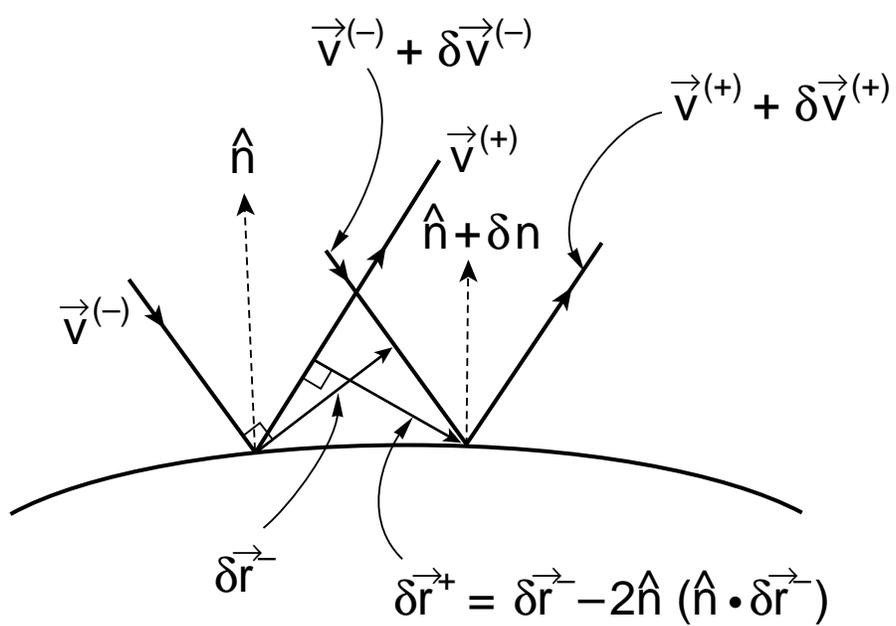
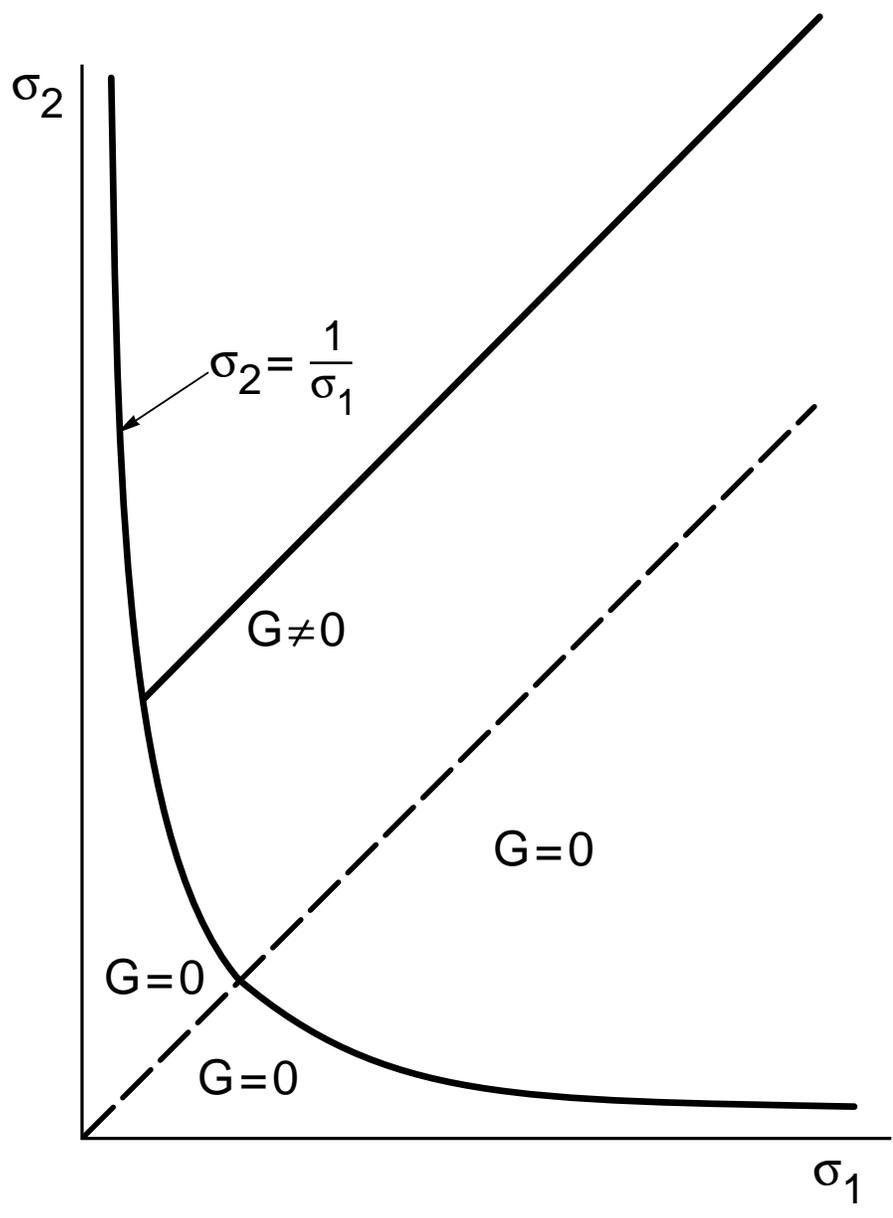
\end{document}